\documentclass[]{aa}  

\usepackage{graphicx}
\usepackage{natbib}
\usepackage{xcolor}
\usepackage{float}
\usepackage{rotating}
\usepackage{booktabs}
\usepackage{mathtools}
\usepackage{placeins}
\usepackage{bm}
\usepackage{longtable}
\usepackage{adjustbox}
\usepackage{supertabular}
\usepackage{multirow}
\usepackage{tabularx}
\usepackage[thinc]{esdiff}
\usepackage{bm}
\usepackage{siunitx}

\usepackage{hyperref}
\hypersetup{
  colorlinks=true,
    linkcolor=black,
    citecolor=blue,
    urlcolor=magenta}

\usepackage{newtxtext}
\usepackage[varvw]{newtxmath}

\begin{document}

   \title{Stochastic excitation of internal gravity waves in rotating late F-type stars: A 3D simulation approach}
   \titlerunning{Gravity waves in rotating late F-type stars}

   \author{S.N.~Breton\inst{1}
          \and  
          A.S.~Brun\inst{2}          
          \and
          R.A.~Garc\'{i}a\inst{2}
          }
    \institute{Universit\'e Paris-Cit\'e, Universit\'e Paris-Saclay, CEA, CNRS, AIM, F-91191, Gif-sur-Yvette, France \\
    \email{sylvain.breton@cea.fr} 
    \and Universit\'e Paris-Saclay, Universit\'e Paris-Cit\'e, CEA, CNRS, AIM, F-91191, Gif-sur-Yvette, France
    }

   \date{}

 \abstract
 {There are no strong constraints placed thus far on the amplitude of internal gravity waves (IGWs) that are stochastically excited in  the radiative interiors of solar-type stars. Late F-type stars have relatively thin convective envelopes with fast convective flows and tend to be fast rotators  compared to solar-type stars of later spectral types. These two elements are expected to directly impact the IGW excitation rates and properties.}   
 {We want to estimate the amplitude of stochastically excited gravity modes (g-modes) in F-type stars for different rotational regimes.}
 {We used the ASH code to perform 3D simulations of deep-shell models of 1.3 $M_\odot$ F-type solar-type stars, including the radiative interior and the shallow convective envelope.} {We found different differential rotation regimes in the convective zone, depending on the rotation rate we imposed on the stellar models. We find that the convective structures and the overshoot properties are affected by rotation. The IGWs are excited by interface interactions between convective plumes and the top of the radiative interior.
 We were able to characterise the IGWs and g-mode properties in the radiative interior, and we compared these properties using the computation from the 1D oscillation code GYRE. 
The amplitude of low-frequency modes is significantly higher in fast-rotating models and the evolution of the period spacing of consecutive modes exhibits evidence of a behaviour that is modified by the influence of the Coriolis force. 
 For our fastest rotating model, we were able to detect the intermediate degree g-mode signature near the top of the simulation domain. Nevertheless, the predicted luminosity perturbations from individual modes still remain at small amplitudes.
 }
 {We obtained mode amplitudes that are several orders of magnitude higher than those of prior 3D simulations of solar models. Our simulations suggest that g-mode signatures could be detectable in late F-type stars, which are the hottest main-sequence solar-type pulsating stars. We therefore emphasise that they constitute object of primary importance for improving our understanding of internal stellar  dynamics.}


   \maketitle

\section{Introduction \label{section:introduction}}

The propagation of internal gravity waves (IGW) in stellar interiors is an expected and well-known phenomenon \citep[see e.g.][and references therein]{ChristensenDaalsgardLectureNotes,Maeder2009,Kippenhahn2012}. These waves are propagative in stably stratified resonant cavities located in radiative regions and evanescent in unstable convective regions. 
Standing IGWs are usually referred as gravity modes (g-modes). In the stellar interior, they may be excited by different mechanisms. In massive stars, small perturbations of the nuclear reaction rate can lead to growing instability in temperature and may therefore excite oscillation modes, an effect known as the $\epsilon$-mechanism \citep[e.g.][]{2012ApJ...749...74M}. In intermediate-mass and massive stars, the $\kappa$-mechanism that is related to opacity-bump in ionization regions is also able to drive mode excitation \citep[e.g.][]{Unno1989,2020FrASS...7...70B}; whereas for early F-type stars pulsating as $\gamma$ Doradus pulsators, a flux blocking mechanism at the bottom of the shallow convective zone is also invoked \citep[e.g.][]{2000ApJ...542L..57G,2005A&A...435..927D}.  
In low-mass stars, IGWs are stochastically excited by turbulent convective motions at the interface between the radiative zone and the convective zone. 
When considering these stochastic mechanisms, the relative importance between the large scale convective plumes penetrating the overshoot regions \citep[e.g.][]{Press1981,1986ApJ...311..563H,1991ApJ...377..268G,1991A&A...252..179Z,1997A&A...322..320Z,2004ApJ...601..512B,2005ApJ...620..432R,2011ApJ...742...79B,2014A&A...565A..42A,2015A&A...581A.112A,2016A&A...588A.122P} and the Reynold stresses in the bulk region above the radiative zone \citep[e.g.][]{1990ApJ...363..694G,2009A&A...494..191B,2010Ap&SS.328..253S,2013MNRAS.430.2363L} has been discussed over the years \citep{2003A&A...405.1025T,2013ApJ...772...21R,2013LNP...865..239P,2015PhRvE..91f3016L,2020ApJ...903...90A,LeSaux2022}.   

Because their properties are intrinsically related to the structure and dynamics of the innermost stellar regions, the question of the g-mode surface amplitude in the Sun and the possibility to observe such modes has therefore been a heavily debated topic since the advent of helioseismology \citep[see e.g.][]{2013ASPC..478..125A}. 
With the introduction of spaceborne asteroseismology and the observation of hundreds of main-sequence solar-type pulsating stars \citep[e.g.][]{2011Sci...332..213C,2022A&A...657A..31M}, it is interesting to consider the case of solar-type pulsators other than the Sun.

Among main-sequence solar-type pulsators, F-type stars are probably the most promising for  g-mode detections. In such stars, fast convective flows \citep[e.g.][]{2017ApJ...836..192B} develop inside a thin surface convective zone, which tend to favour the tunneling of g modes towards the surface. Most of observed F-type stars also tend to be fast rotators: in the \textit{Kepler} \citep{Borucki2010} catalog provided by \citet{2021ApJS..255...17S}, the median surface rotation period for F-type stars with measured rotation rate is close to 6 days. Through the action of the Coriolis force, rotation is expected to have an effect on IGW behaviour \citep{1997ApJ...491..839L,2003MNRAS.340.1020T} by turning them into gravito-inertial waves, which remain propagative in an unstable layer if their frequency, $\omega,$ is below twice the local rotation frequency \citep{2014A&A...565A..47M}. In recent years, the asteroseismology of fast rotators has presented spectacular developments both with regard to theoretical \citep[e.g.][]{2019A&A...627A..64P,2020A&A...636A.100P,2021A&A...656A.122D,Dhouib2021,Dhouib2022} and observational \citep[e.g.][]{2015ApJS..218...27V,2016A&A...593A.120V,2017A&A...598A..74P,2021MNRAS.503.5894S} aspects, which has enabled probes of the internal dynamics of intermediate-mass and massive stars at the interface between the convective core and the radiative interior \citep[e.g.][]{2017MNRAS.465.2294O,2019A&A...626A.121O,Ouazzani2020,2020MNRAS.491.3586L,2021A&A...656A.121A,2021NatAs...5..715P}. 
Stochastically excited gravito-inertial modes have also been detected in rapidly rotating massive stars \citep{Neiner2012,Neiner2020}.
Due to the difficulties in observing g-modes in solar-type pulsators discussed above, it appears that performing 3D deep-shell simulations of late F-type stars constitutes a first approach to understanding the interplay between IGWs and rotation in their radiative interiors. While 3D simulations of F-type stars have already been performed, they have only been aimed at studying the convective-envelope dynamics \citep{Augustson2012} or dynamo properties \citep{Augustson2013}.     

To this day, the range of stellar masses explored by IGW 3D hydrodynamic simulations remains limited. \citet{2011ApJ...742...79B} and \citet[referred as A14 and A15, respectively]{2014A&A...565A..42A,2015A&A...581A.112A}, performed solar-model simulations to evaluate the possibility to detect individual g-modes with helioseismic instruments.
\citet{2004ApJ...601..512B} presented a simulation of the central regions of a 2 M$_\odot$ A star, where IGWs are excited by convective motions in the core.
\citet{2019ApJ...876....4E} extended the 2D simulations wave excitation analysis of 3 M$_\odot$ stars
from \citet{2013ApJ...772...21R} into the 3D space. 
\citet{Augustson2016} studied the core-radiative interior interplay in  10 M$_\odot$ O stars, while \citet{Andre2019} examined the breaking of waves in 15 M$_\odot$ O stars.
In this work, we intend to extend the study of IGWs in low-mass stars to non-solar models with varying rotation rates. We present the first deep-shell hydrodynamical simulations of rotating F-type stars. The layout of the paper is as follows. In Sect.~\ref{sec:numerical_model}, we present the model and cases for which we solve the hydrodynamical equations.
In Sect.~\ref{sec:fstar_dynamics}, we describe the global dynamics of our F-type star model for the different rotation regimes we considered.
The IGWs and g-mode properties are extensively described and analysed in Sect.~\ref{sec:igw}. In particular, we estimate the evolution with frequency of rotational splittings and period spacing. We then estimate the surface mode visibility of g modes in Sect.~\ref{sec:surface_visibility}. We discuss the theoretical and observational perspectives opened up by this work in Sect.~\ref{section:conclusion}.

\section{Numerical setup \label{sec:numerical_model}}

\subsection{Model equations \label{sec:equations}}

We used the ASH code \citep{CLUNE1999361,2004ApJ...614.1073B} to solve the 3D hydrodynamic equations in the anelastic approximation. We consider a system of spherical coordinates $(r, \theta, \phi)$ in a frame rotating at constant angular velocity $\bm{\Omega}_0 = \Omega_0 \bm{e}_z$.
The reference density, pressure, temperature, and specific entropy are denoted as $\bar{\rho}$, $\bar{P}$, $\bar{T}$, and $\bar{S}$, while the fluctuations about this reference state are $\rho$, $P$, $T$, $S$. 
Following the prescription from \citet{2012ApJ...756..109B} and using the implementation presented in A14, we used the Lantz-Braginsky-Roberts \citep[][LBR]{1992PhDT........78L,1995GApFD..79....1B} equations to define the momentum equation. 
Indeed, the wave energy may be overestimated when using the traditional anelastic approximation.
In the LBR formulation, the reduced pressure $\tilde{\omega}=P/\bar{\rho}$ is considered instead of the fluctuating pressure $P$. The non-linear momentum equation is therefore:
\begin{equation}
\label{eq:momentum}
\bar{\rho} \left( \diffp{\bm{v}}{t} + ({\bm{v}.\nabla}) \bm{v} \right)
= - \bar{\rho} \nabla \tilde{\omega} - \bar{\rho} \frac{S}{c_p} \bm{g} - 2\bar{\rho}\bm{\Omega}_0 \times \bm{v} - \nabla.\bm{\mathcal{D}} \; ,
\end{equation}
where $\bm{v} = (v_r, v_\theta, v_\phi)$ is the local velocity, $\bm{g}$ is the gravitational acceleration, and $\bm{\mathcal{D}}$ is the viscous stress tensor:
\begin{equation}
    \label{eq:reynold_tensor}
    \mathcal{D}_{ij} = - 2 \bar{\rho} v \left( e_{ij} - \frac{1}{3}(\nabla.\bm{v}) \delta_{ij} \right) \; ,
\end{equation}
with $e_{ij} = 1/2 \left( \partial_j v_i + \partial_i v_j \right)$ the strain rate tensor and $\delta_{ij}$ the Kronecker symbol. In the anelastic approximation, the continuity equation is expressed as:
\begin{equation}
    \nabla . (\bar{\rho} \bm{v}) = 0.
\end{equation}

We assume a linearised equation of state and the zeroth-order ideal gas law:
\begin{equation}
    \label{eq:eq_of_state}
    \frac{\rho}{\bar{\rho}} =  \frac{P}{\bar{P}} -  \frac{T}{\bar{T}}
    =  \frac{P}{\gamma \bar{P}} - \frac{S}{c_p} \; ,
\end{equation}
\begin{equation}
    \bar{P} = \mathcal{R} \bar{\rho} \bar{T} \; ,
\end{equation}
where $\gamma$ is the adiabatic exponent, $c_p$ is the specific heat per unit mass at constant pressure, and $\mathcal{R}$ is the gas constant. Finally, the equation of conservation of internal energy is:
\begin{multline}
    \label{eq:conservation_energy}
    \bar{\rho} \bar{T} \diffp{S}{t} + \bar{\rho}\bar{T}\bm{v}.\nabla \Big( S + \bar{S} \Big) 
    = \bar{\rho}\epsilon + \nabla. \Bigg[ \kappa_r\bar{\rho} c_p \nabla \Big( T + \bar{T} \Big) \\
    + \kappa \bar{\rho} \bar{T} \nabla S + \kappa_0 \bar{\rho} \bar{T} \nabla \bar{S}  \Bigg] 
    + 2 \bar{\rho} \nu \left[ e_{ij}e_{ij} - \frac{1}{3} (\nabla.\bm{v})^2 \right] \; ,
\end{multline}
where $\kappa_r$ is the radiative diffusivity and $\bar{\rho}\epsilon$ the volume-heating term related to the energy generation by nuclear burning. The $\epsilon$ profile is parametrised as a power law of $\bar{T}$, $\epsilon = \epsilon_0 \bar{T}^k$. The parameters $\epsilon_0$ and $k$ are computed in order to have the integrated heating equal to the stellar luminosity at the top of the radiative zone.  
The parameters $\nu$ and $\kappa$ are the effective diffusivities representing the unresolved momentum and heat-transport by subgrid-scale (SGS) motions, while the diffusivity, $\kappa_0$, is set to carry the unresolved entropy eddy flux in the convective zone near the surface. We ensure that this flux does not play any role in the radiative zone by chosing a $\kappa_0$ profile that decreases exponentially with depth \citep{2000ApJ...532..593M}.  

\begin{figure}[ht!]
    \centering
    \includegraphics[width=0.48\textwidth]{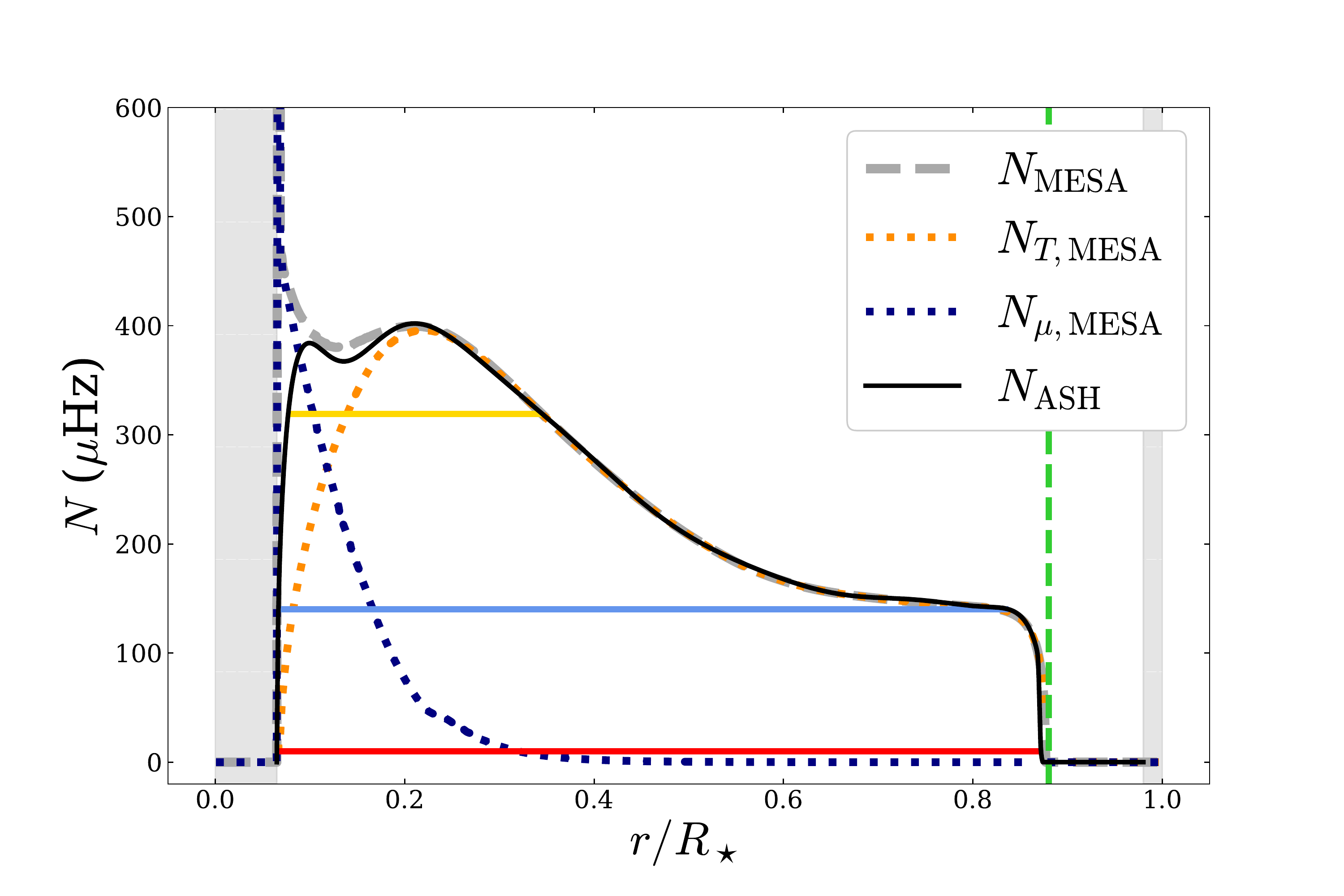}
    \caption{Radial profiles of the Brunt-Väisälä frequency $N$ in the ASH model (black) and the MESA model (dashed grey). The structural and chemical contributions to $N_\mathrm{MESA}$, $N_{T,\mathrm{MESA}}$, and $N_{\mu,\mathrm{MESA}}$, are represented in dotted-orange and dotted-dark-blue lines, respectively. The dashed green vertical line shows the boundary between the radiative zone and the convective envelope, while the grey hatched regions are excluded from the simulation domain. The frequency corresponding to the ray paths of Fig.~\ref{fig:raytracing} are represented by the solid horizontal red, blue, and yellow lines.}
    \label{fig:bv}
\end{figure}

\begin{figure}[ht!]
    \centering
    \includegraphics[width=0.48\textwidth]{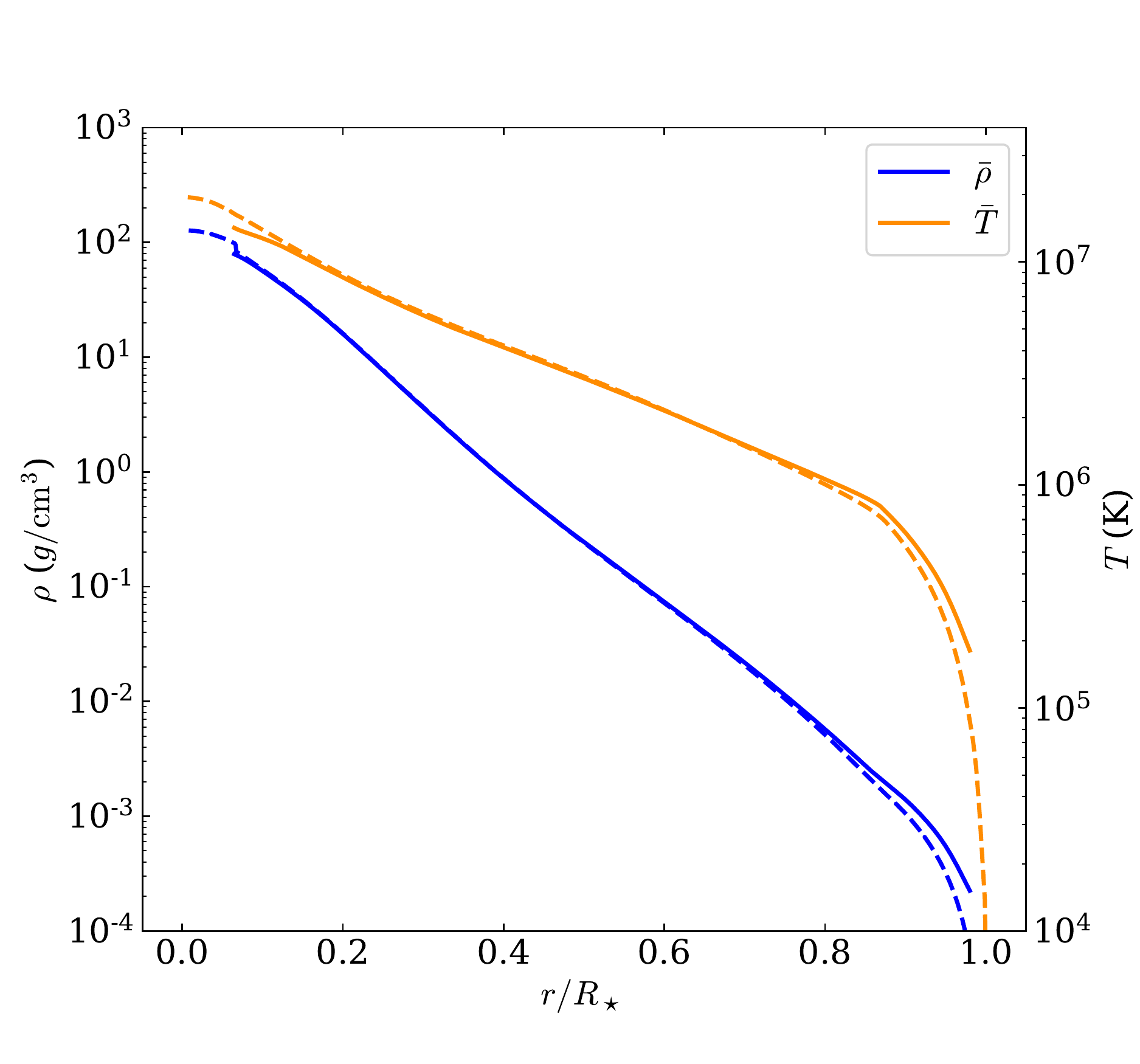}
    \caption{Density $\bar{\rho}$ (blue) and temperature $\bar{T}$ (orange) profiles used as reference profiles for the simulation are represented with straight lines. MESA model profiles are shown in dashed lines for comparison.}
    \label{fig:rho_T}
\end{figure}

\subsection{Models \label{sec:models}}

\begin{figure}[ht!]
    \centering
    \includegraphics[width=0.48\textwidth]{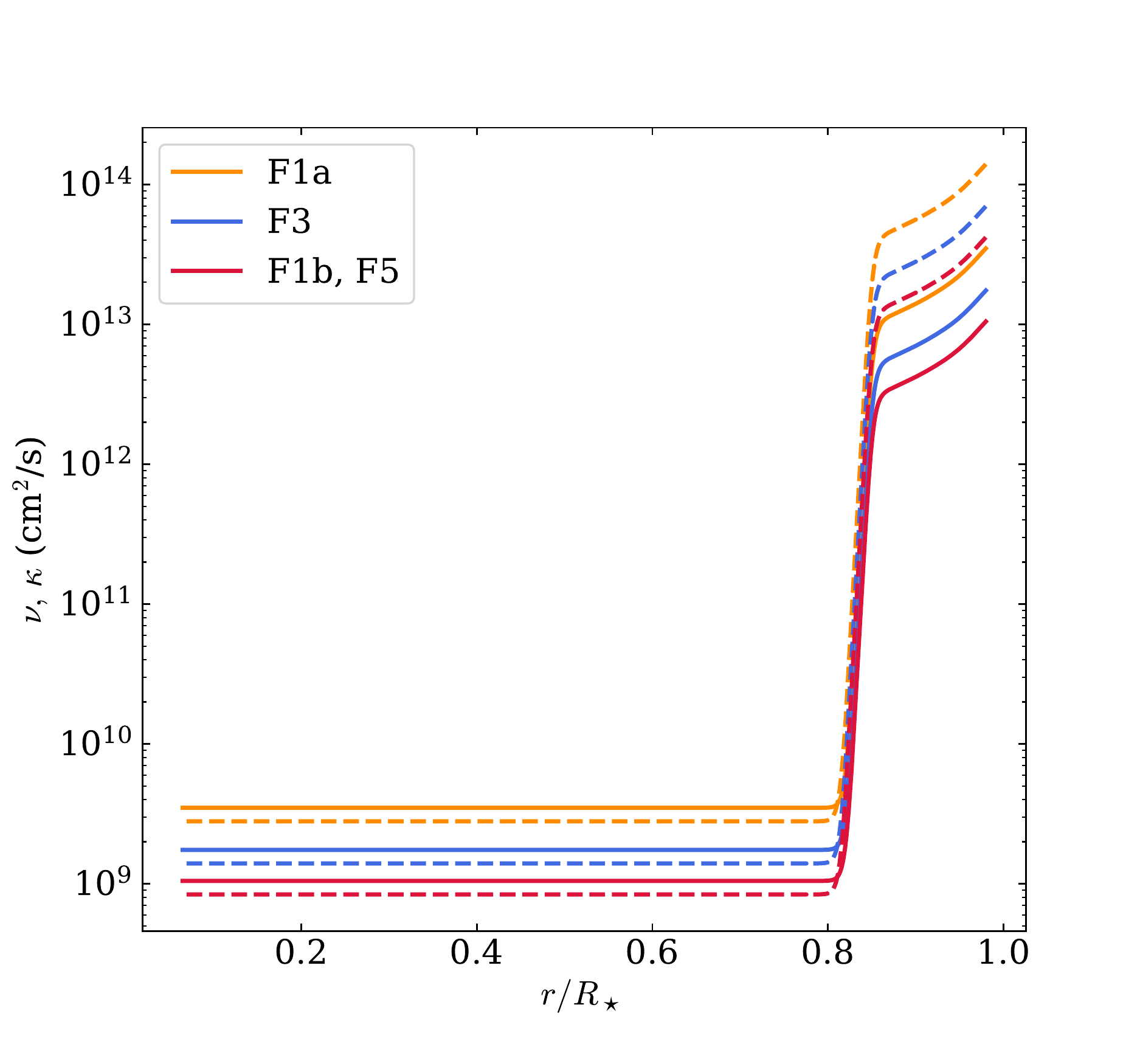}
    \caption{Radial profile of the diffusivities, $\nu$ (\textit{straight lines}) and $\kappa$ (\textit{dashed lines}), for the F1a (orange), F3 (blue), F1b, and F5 (red) cases.}
    \label{fig:diffusivity}
\end{figure}

We used the Modules for Experiments in Stellar Astrophysics
\citep[MESA,][]{Paxton2011, Paxton2013, Paxton2015, Paxton2018, Paxton2019} to generate a 1D model of a $1.3$ M$_\odot$ star. We considered the model at the evolutionary stage when its hydrogen center mass fraction is 0.35. At this stage, the model luminosity, $L_\star$, is $3.31 \, \rm L_\odot$, the effective temperature, $T_\mathrm{eff}$, is 6353 K and the logarithm of the surface gravity, $\log g,$ is 4.2.
The stellar radius $R_\star$ is $1.0465\times10^{11}$ cm (1.5 R$_\odot$) and for the simulation, we considered the region spanning from 0.07 to 0.98~$R_\star$ (purposely omitting the relatively small convective core). We denote $r_\mathrm{bottom}$ and $r_\mathrm{top}$ the innermost and outermost radius of our simulation region, respectively.
The upper edge of the convective core is treated as an impenetrable boundary and we consider the following boundary conditions at the top and bottom of the domain:
\begin{enumerate}
    \item rigid: $v_r|_{r_\mathrm{top}} = v_r|_{r_\mathrm{bottom}} = 0$ ;
    \item stress-free: $\diffp{}{r}\left(\frac{v_\theta}{r}\right)\bigg|_{r_\mathrm{top}, \, r_\mathrm{bottom}} = \diffp{}{r}\left(\frac{v_\phi}{r}\right)\bigg|_{r_\mathrm{top}, \, r_\mathrm{bottom}} = 0$ ;
    \item constant mean entropy gradient: \\
    $\diffp{\bar{S}}{r}\bigg|_{r_\mathrm{bottom}} = 0$, \\
    $\diffp{\bar{S}}{r}\bigg|_{r_\mathrm{top}} = -\num{1.5e-6}$ cm.K$^{-1}$.s$^{-2}$.
\end{enumerate}

The interface between the radiative zone and the upper convective zone is located at $r_\mathrm{CZ} = 0.88$~$R_\star$ (\num{9.18e10} cm). We note that the convective zone absolute thickness of our model is approximately 60\% the thickness of the solar convective zone, while the total volume of the convective shell is approximately 68\% larger than for the solar case.  
Concerning the energy generation parameter, we performed a fit on the MESA luminosity profile to obtain $\epsilon _0=3.23\times10^{-7}$ and $k=7.414$. The $\mathrm{d}S/\mathrm{d}r$ and $g$ profiles provided to ASH are obtained from a polynomial fit of the MESA model. These two profiles yield the Brunt-Väisäla frequency profile $N$ (given in Hz) according to the relation \citep[e.g.][]{Maeder2009}
:\begin{equation}
    \label{eq:brunt_vaisala}
    N^2 (r) = \frac{1}{2 \pi}\frac{g}{c_p} \diff{S}{r} \; , 
\end{equation}
where we take a uniform $c_p = \num{3.42e8}$ erg.g$^{-1}$.K$^{-1}$ for our input ASH profile. We know that IGWs are propagative in regions where $N^2>0$ and evanescent in regions where $N^2<0$.
Close to the convective core, the $\mathrm{d}S/\mathrm{d}r$ and $N$ profiles are affected by the chemical gradient as it can be seen in Fig.~\ref{fig:bv}, where we show the structural contribution, $N_{t, \mathrm{MESA}}$, and the chemical contribution, $N_{\mu, \mathrm{MESA}}$, to the MESA Brunt-Väisälä profile, $N_\mathrm{MESA}$. 
The extent of the convective envelope and the areas from the MESA models excluded from the 3D simulation domain are represented in the figure as well. 
We also compared $N_\mathrm{MESA}$ to the $N_\mathrm{ASH}$ profile obtained from the $\mathrm{d}S/\mathrm{d}r$ and $g$ fits. Although we are not able to reproduce the stiff frequency bump at the bottom of the radiative zone, we find only a 3.7\% discrepancy when we integrate $N_\mathrm{ASH}$ and $N_\mathrm{MESA}$ along $r$. When we compute the $\int_{r_\mathrm{bottom}}^{r_\mathrm{top}} N/r \, \mathrm{d} r$, which is directly related to the g-mode period spacing, we find a 11.6\% discrepancy, which means that the g-mode properties obtained from the 1D MESA model cannot be  compared to the 3D simulations in a straightforward way. The trends and global properties could certainly be compared between the two set of profiles but in order to be as consistent as possible, we will compare the outcome of the nonlinear 3D simulations with the g-mode frequencies predicted by a 1D oscillation code using the 1D ASH reference profile (see Fig.~\ref{fig:E_l_depth_80}) and Section~\ref{sec:gyre_comparison}. 
The $\bar{\rho}$ and $\bar{T}$ are computed using a Newton-Raphson algorithm in order to set the reference state at the hydrostatic equilibrium. In Fig.~\ref{fig:rho_T}, we compare the profiles obtained with the Newton-Raphson algorithm and the reference profiles from MESA. Despite small deviations in the convective zone, we note that there is a good overall agreement. 

\begin{figure}[ht!]
    \centering
    \includegraphics[width=.48\textwidth]{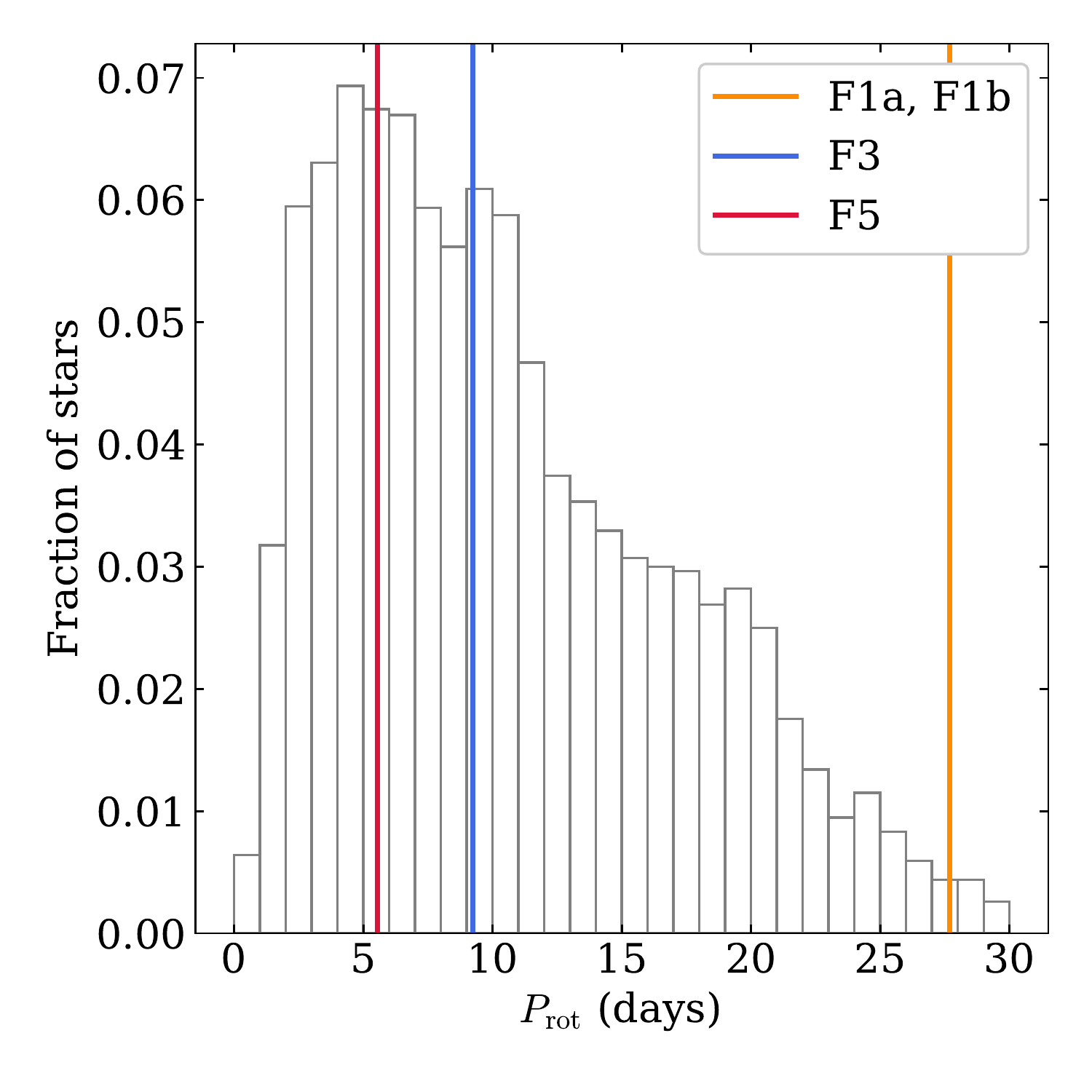}
    \caption{Distribution  of surface rotation periods $P_\mathrm{rot}$ measured by \citet{2021ApJS..255...17S} in the \textit{Kepler} data, considering stars with $6000 < T_\mathrm{eff} < 6600$~K and $\log g > 4$~dex \citep[using values from][]{Berger2020}. Vertical colour lines indicate the periods corresponding to the rotation rates selected for our simulations.}
    \label{fig:santos_2021_distribution}
\end{figure}

To study the impact of rotation on the model differential rotation in the convective zone and on gravity waves dynamics in the radiative interior, we ran cases with $\Omega_0=1$ (F1a and F1b cases), 3 (F3 case), and 5~$\Omega_\odot$ (F5 case), where we take $\Omega_\odot = \num{2.6e-6}$ rad.s$^{-1}$. All the considered cases verify that $\Omega_0 \ll \Omega_K = \sqrt{\mathrm{G}M_\star/R_\star^3} \sim \num{3.9e-4}$ rad.s$^{-1}$, with $\Omega_K$ the Keplerian critical breakup angular velocity and $\mathrm{G}$ the universal gravitational constant. Indeed, 5 $\Omega_\odot$ corresponds to $\sim 3.4$~\% of $\Omega_K$ for the considered stellar model. The centrifugal deformation of the star can therefore be neglected \citep[see][for a discussion]{Zahn1992}. 
We also emphasise the fact that, for this range of rotation rate, the relative difference in radius is below 0.1\%, and that the position of the convective core differs by no more than 1\% of the relative radius \citep[][and Amard, private communication]{Amard2019}, justifying our choice to use the same reference structure for all cases.
As shown in Fig.~\ref{fig:santos_2021_distribution}, where we represent the distribution of photometric surface rotation periods from the \citet{2021ApJS..255...17S} \textit{Kepler} catalog for stars in the range $6000 < T_\mathrm{eff} < 6600$~K and $\log g > 4$~dex \citep[using the values from][]{Berger2020}, the F1a and F1b cases are close to the slow-rotator tail of the distribution for this population of stars.  
We emphasis that the F3 case is close to the median of the distribution (9.4 days) and the F5 case to the distribution maximum. Our choice of rotation rates is therefore a good representation of what is observed in the \textit{Kepler} sample.

In order to simulate both the convective and the radiative zone, we take the diffusivities, $\nu$ and $\kappa,$ as function of the radius and we use the following profiles:
\begin{align}
    \label{eq:diffusivity}
    \nu (r) &= \nu_\mathrm{top} \Bigg[ \beta_\nu + \frac{1 -\beta_\nu}{2} \left( \frac{\bar{\rho}_\mathrm{top}}{\bar{\rho}}  \right)^\frac{1}{2} \left(\tanh \frac{r-r_t}{\sigma_t} + 1\right) \Bigg] \;, \\
    \kappa (r) &= \kappa_\mathrm{top} \Bigg[ \beta_\kappa + \frac{1-\beta_\kappa}{2} \left( \frac{\bar{\rho}_\mathrm{top}}{\bar{\rho}}  \right)^\frac{1}{2} \left(\tanh \frac{r-r_t}{\sigma_t} + 1\right) \Bigg] \;,
\end{align}
with $\bar{\rho}_\mathrm{top} = \bar{\rho}(r_\mathrm{top})$, $\bar{\nu}_\mathrm{top} = \bar{\nu}(r_\mathrm{top})$, $\bar{\kappa}_\mathrm{top} = \bar{\kappa}(r_\mathrm{top})$, $\sigma_t = 0.08$ the profile stiffness parameter, $\beta_\nu = 10^{-4}$, and $\beta_\kappa = 2\times10^{-5}$. The Prandtl number, $Pr = \nu / \kappa$, is therefore not uniform along the stellar radius, from 0.25 at the top of the domain to 1.25 at the bottom. We show in Fig.~\ref{fig:diffusivity}, the diffusivity profiles adopted for the F1a, F1b, F3, and F5 cases. 
We emphasise that, with this choice of profile, we are able to resolve the motions in the convective envelope, while the abrupt drop of $\nu$ and $\kappa$ in the tachocline limits the viscous and radiative damping of the IGWs in the radiative interior.
We remind here that 3D stellar models must be compared with each other with the required level of caution. Indeed, as numerical limitations of simulations prevent us from reaching actual stellar regimes, it is important to take into account and discuss the role of the different characteristic fluid numbers exhibited by the different cases. Supercriticality considerations related to the choice of $\nu_\mathrm{top}$ and $\kappa_\mathrm{top}$ will be discussed in Section~\ref{sec:rotation_and_convection} while the influence of the Rossby number $Ro$ on the dynamic of the convective envelope will be exposed in Section~\ref{sec:rot_diff}.
The properties of the four different simulations are summarised in Table~\ref{tab:model_summary}.

\subsection{Numerical resolution}

\begin{figure}[ht!]
    \centering
    \includegraphics[width=0.48\textwidth]{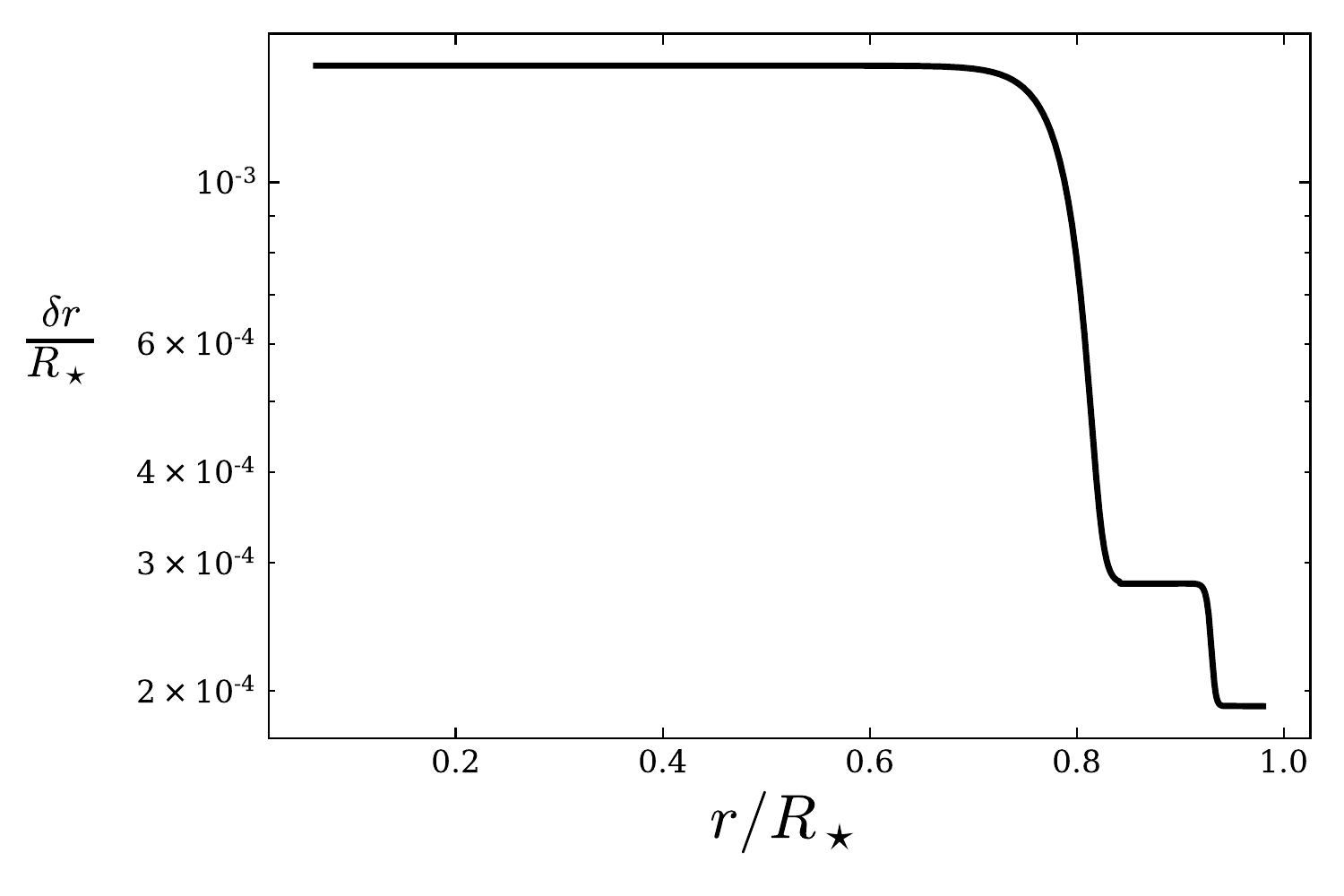}
    \caption{Spacing in the non-uniform radial grid used in the 3D ASH simulations.}
    \label{fig:grid_resolution}
\end{figure}

In order to solve the hydrodynamic equations, following A14, the horizontal structure of the velocity and thermodynamic variables are expanded in spherical harmonics $Y_{\ell,m} (\theta, \phi)$, with $\ell$ the spherical degree and $m$ the azimuthal number (see Table~\ref{tab:model_summary} for $N_\theta \times N_\phi$ resolution),  while for the radial structure we use a non-uniform-grid with a finite difference approach. The grid we use has $N_r = 1205$ radial points. The radiative zone is resolved with 764 points while the convective zone is resolved with 441 points.
As shown in Fig.~\ref{fig:grid_resolution}, the radial resolution is significantly finer in the convective zone and in the overshoot region, where the diffusivity values, $\nu$ and $\kappa,$ drop abruptly. 
The equations are solved using an explicit Adams-Bashforth time integration scheme for the advection and Coriolis terms, while the diffusive and buoyancy terms are treated through a semi-implicit Crank-Nicolson method \citep{1984JCoPh..55..461G,CLUNE1999361}.

\begin{figure}[ht!]
    \centering
    \includegraphics[width=0.48\textwidth]{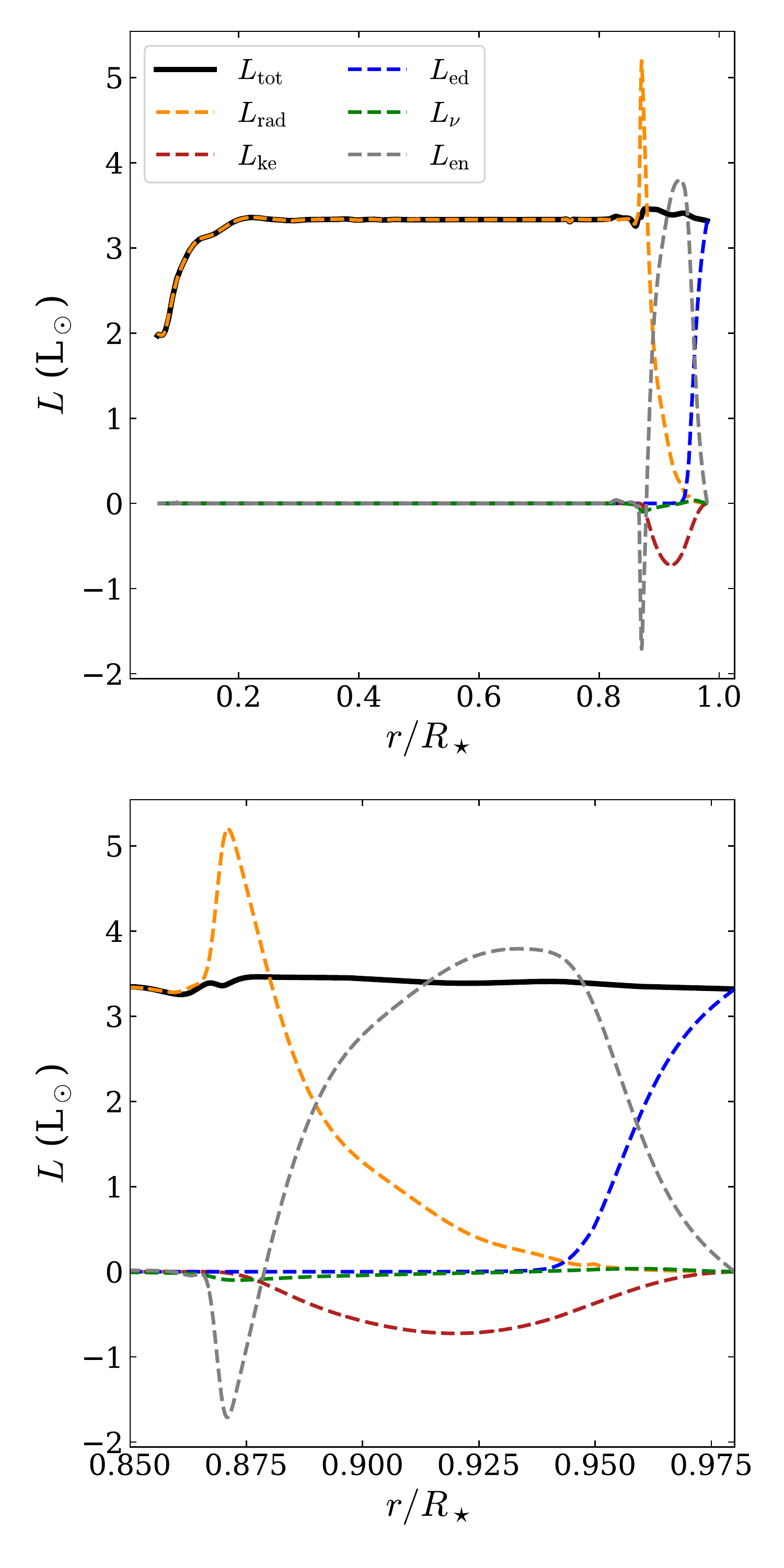}
    \caption{Flux balance between radiative luminosity $L_\mathrm{rad}$ (dashed orange), kinetic energy luminosity $L_\mathrm{ke}$ (dashed red), enthalpy luminosity $L_\mathrm{en}$ (dashed grey), diffusivity luminosity $L_\nu$ (dashed green), and subgrid-scale eddy luminosity $L_\mathrm{ed}$ (dashed blue) for the F5 case. The total luminosity, $L_\mathrm{tot}$, is represented in black. The top panel shows the flux balance for the whole simulation domain, while the bottom panel shows an enlargement in the convective zone and the overshoot region.}
    \label{fig:flux_balance}
\end{figure}

Once the simulation has been evolved over several tens of convective overturning times ($4 < \tau_\mathrm{conv} < 5$~days in our simulations), we obtain the flux balance between the different energy transport processes represented in Fig.~\ref{fig:flux_balance}. In our hydrodynamic setup, the total luminosity, $L_\mathrm{tot}$, can be decomposed as 
\begin{equation}
    L_\mathrm{tot} = L_\mathrm{rad} + L_\mathrm{ke} + L_\nu + L_\mathrm{en} + L_\mathrm{ed} \; ,
\end{equation}
where $L_\mathrm{rad}$ is the radiative flux, $L_\mathrm{ke}$ the kinetic energy flux, $L_\nu$ the diffusive processes energy flux, $L_\mathrm{en}$ the enthalpy flux, and $L_\mathrm{ed}$ the unresolved eddy flux (as described in Sect.~\ref{sec:equations}). 
As expected, the energy transport is purely radiative below the tachocline. 
 The unresolved energy flux becomes dominant near the top of the simulation domain (as explained in Sect.~\ref{sec:equations}).
In the middle of the convective zone, the enthalpy flux excess compensates for the inwards kinetic energy flux. 
At the interface between the radiative and the convective zone, we note that a significant amount of enthalpy is transported towards the interior due to overshoot mechanisms in the tachocline. 
As the timescales we consider in the simulation are much smaller than the thermal relaxation timescale (about \num{1e5} yr) required for the system to reach a new equilibrium \citep{1991A&A...252..179Z}, we modified the $\kappa_r$ value at the interface in order to increase the radiative flux and balance the enthalpy flux excess in this region \citep{2000ApJ...532..593M,2011ApJ...742...79B}, thus easing the relaxation time. 

\begin{table*}[]
    \centering
    \caption{Global properties for the considered cases.}
    \begin{tabular}{ccccc}
    \hline \hline
    Case & F1a & F1b & F3 & F5 \\
    \hline
    ($N_r$, $N_\theta$, $N_\phi$) & (1205, 512, 1024) & (1205, 1024, 2048) & (1205, 1024, 2048) & (1205, 1024, 2048) \\
    $\Omega_0$ ($\Omega_\odot$) & 1 & 1 & 3 & 5 \\
    $\nu_\mathrm{top}$ (cm$^2$/s) & \num{3.5e13} & \num{1.05e13} & \num{1.75e13} & \num{1.05e13} \\
    $\kappa_\mathrm{top}$ (cm$^2$/s) & \num{1.4e14} & \num{4.2e13} & \num{7e13} & \num{4.2e13} \\
    $r_t$ (cm) & \num{8.9e10} & \num{8.9e10} & \num{8.9e10} & \num{8.9e10} \\
    Longest time series (day) & 117 & 52 & 48 & 130 (280) \\
    \end{tabular}
    \tablefoot{
    The last row of the table specifies the longest $v_r$ time series we have for each case (with a time sampling $dt$ below 1250s to keep the Nyquist frequency of the series above $N_\mathrm{max} = 400$~$\mu$Hz, see Sect.~\ref{sec:igw_spectrum}.), with outputs saved for one depth every 1\% of the stellar radius. The value between parenthesis in the F5 case correspond to the length of the time series for which we save outputs only at $r = 0.13 \, R_\star$. This time series is used in Sects.~\ref{sec:period_spacing} and \ref{sec:splittings}.  
    }
    \label{tab:model_summary}
\end{table*}

\section{F-type star dynamics in 3D \label{sec:fstar_dynamics}}

\begin{figure}[ht!]
    \centering
    \includegraphics[width=0.48\textwidth]{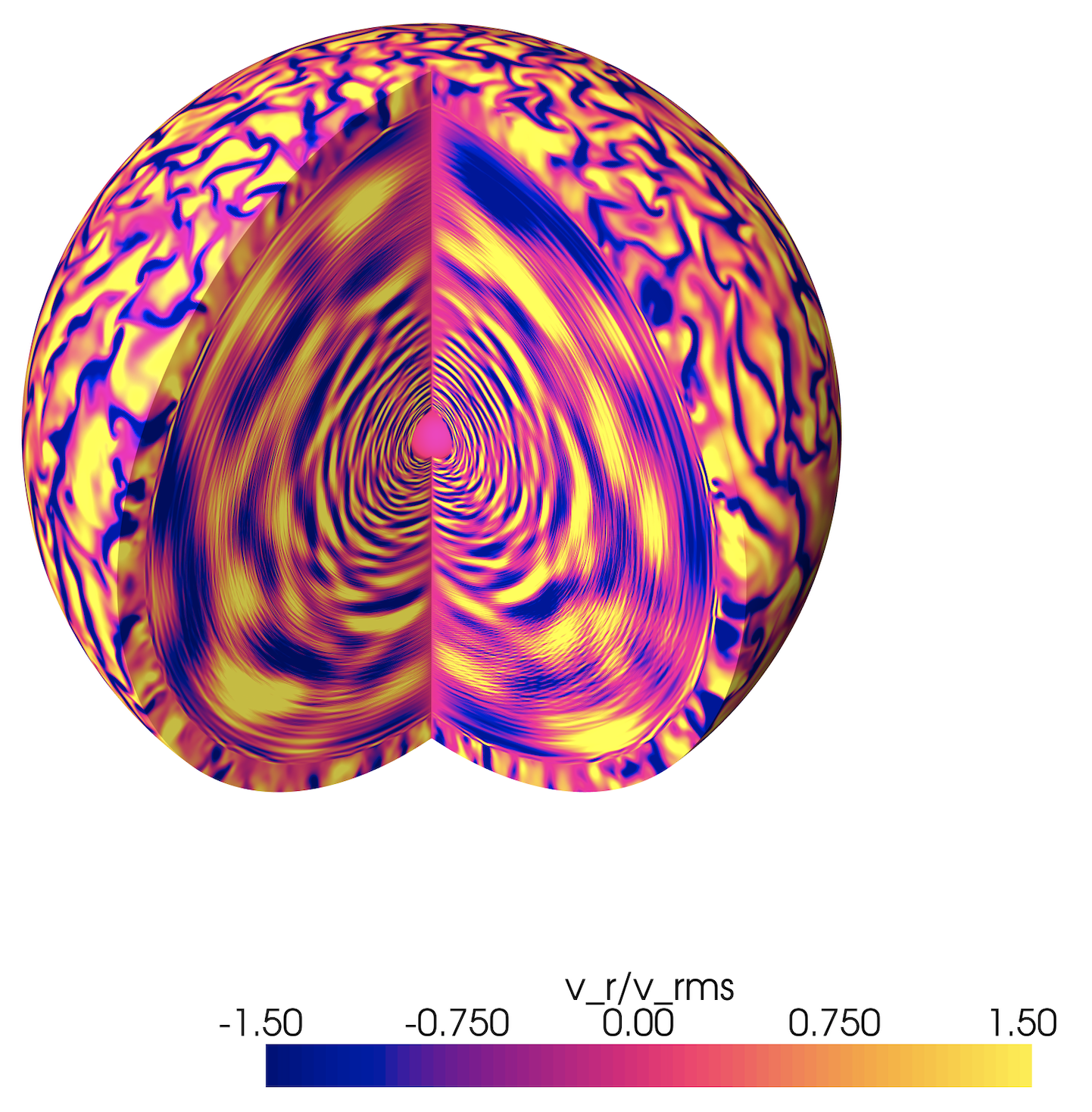}
    \caption{3D volume of the radial velocity, $v_r$, normalised by the rms velocity, $\tilde{v}_r$ (denoted \texttt{v\_rms in the colorbar)} for the F5 case, from $r = 0.07$ to 0.94 $R_\star$. Upward flows are in represented in yellow and downward flows in blue.}
    \label{fig:3d_volume}
\end{figure}

To illustrate the different behaviours of the radiative zone and the convective zone, we represent a 3D volume of the F5 case in Fig.~\ref{fig:3d_volume}.
We show the radial velocity, $v_r$, normalised by its root-mean square (rms) value at each given radius, $\tilde{v}_r$.
In the convective zone, large convective structures are shaped by the stratification of the reference state, with wide upward flows and thinner downwards flows. These downflow plumes act as an excitation piston when they interact with the stratified radiative zone. Below the interface, excited IGWs propagate towards the convective core and are reflected at the bottom of the radiative zone. We note the change of convection scales as we reach the base of the convective zone and the presence of concentric rings typical of low-frequency IGWs (see Section~\ref{sec:igw}). 
We must draw attention to the fact that, even if the hydrodynamical equations we solve are deterministic, everything occurs as if the IGWs are stochastically excited, since the properties of the plumes interacting at any given moment with the top of the radiative zone are not known a priori.

\subsection{Rotation and convection \label{sec:rotation_and_convection}}

\begin{figure}[ht!]
    \centering
    \includegraphics[width=0.49\textwidth]{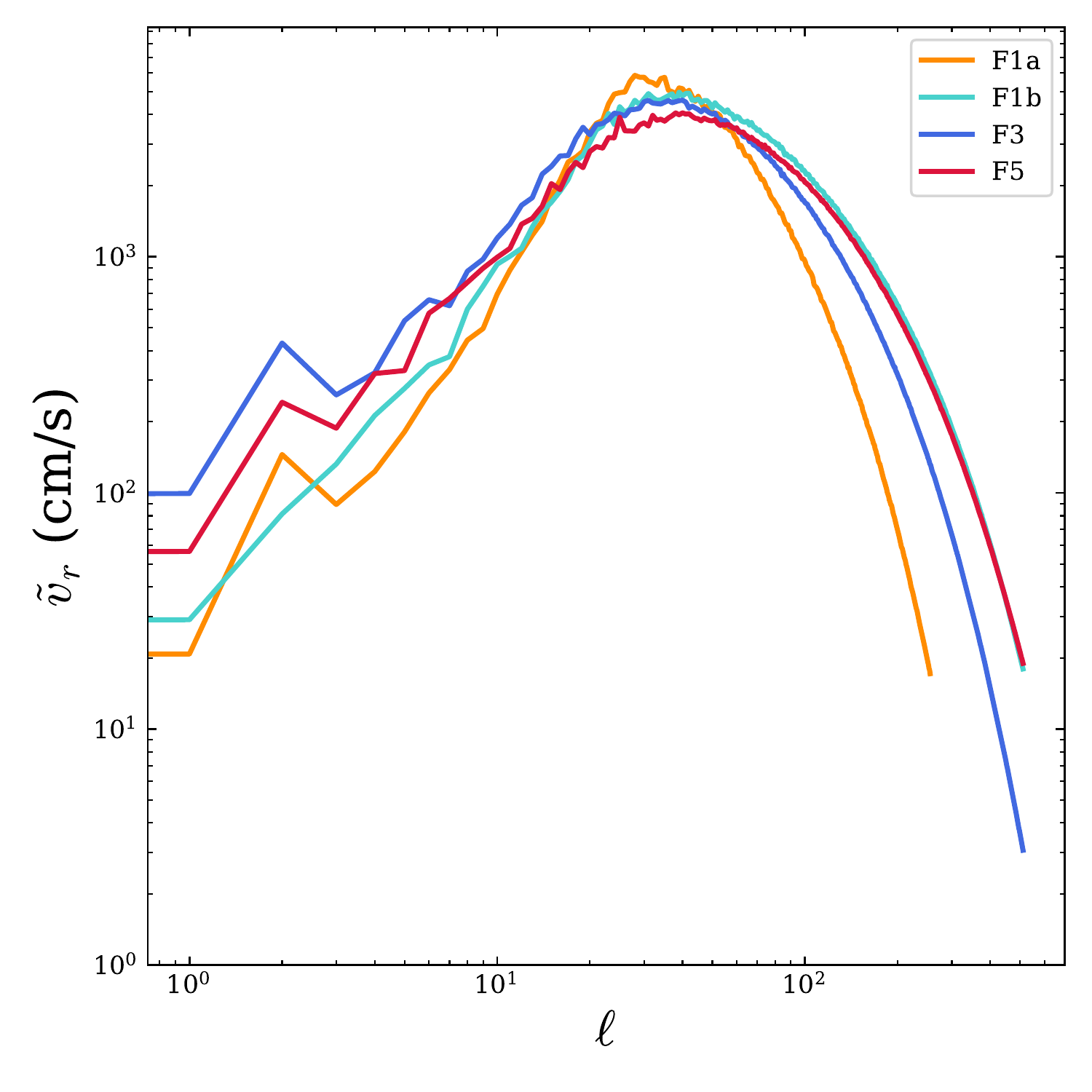}
    \includegraphics[width=0.49\textwidth]{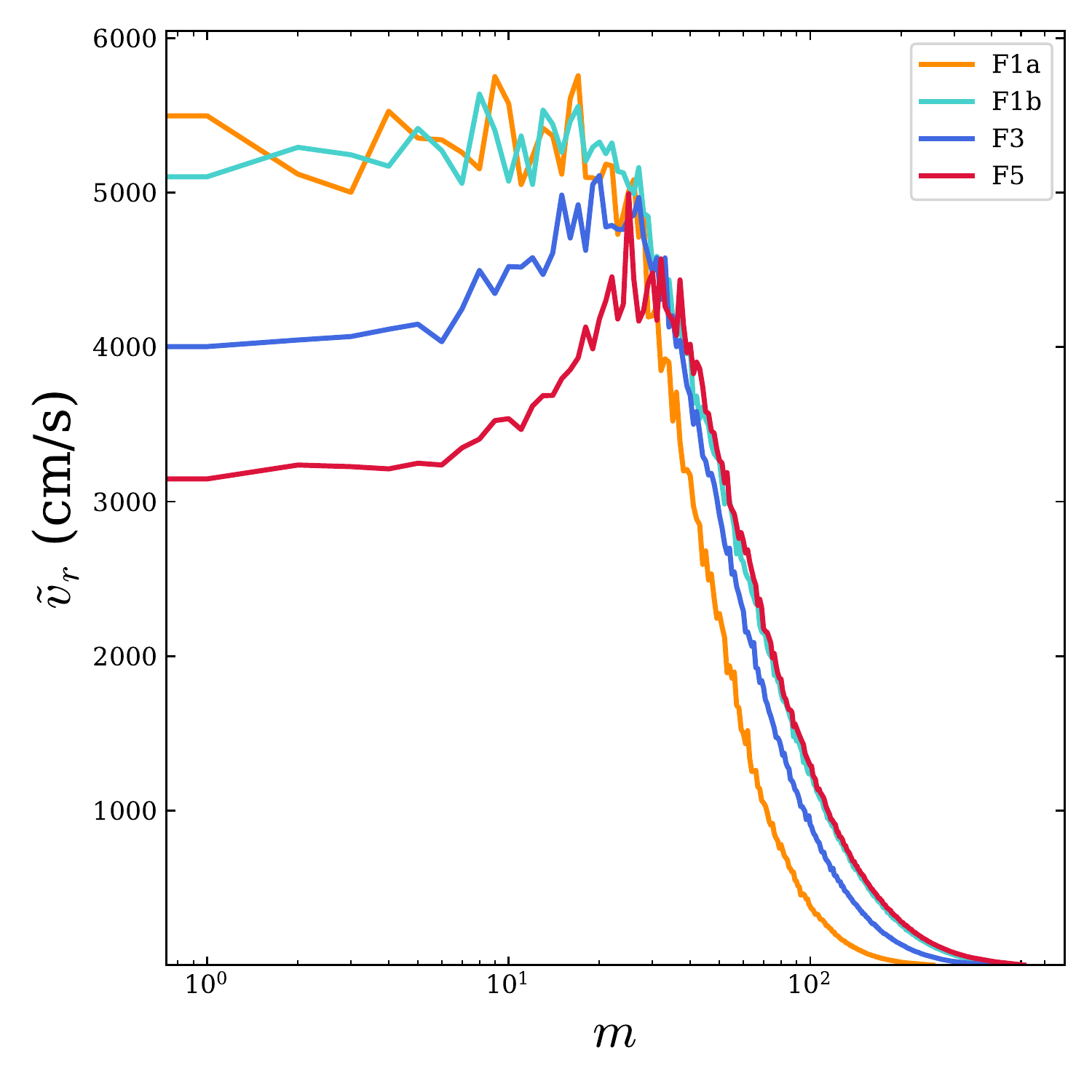}
    \caption{Spherical harmonic decomposition of the time average of the rms radial velocity in the middle of the convective zone ($r = 0.93$ $R_\star$) for the F1a (orange), F1b (cyan), F3 (blue), and F5 (red) cases. In the top panel, the rms components are summed over $m$ and shown as a function of $\ell$. In the bottom panel, the rms components are summed over $\ell$ and shown as a function of $m$.}
    \label{fig:convection_spectrum}
\end{figure}

In order to assess the degree of turbulence of the flows in the convective zone, we estimate the Reynolds number, $Re$, as:
\begin{equation}
    Re = \frac{\tilde{v}_r L}{\nu} \; ,
\end{equation}
where $\tilde{v}_r$ is the root-mean-square (rms) radial velocity and $L$ a characteristic length. We consider $L = d_\mathrm{CZ}$, where $d_\mathrm{CZ}$ is the thickness of the convective zone. As expected from the chosen $\nu$ profiles, the most turbulent case is the F1b case, with a rms $Re$ of 73 in the middle of the convective zone, followed by the F5 case with $Re = 64$. The F1a and F3 cases have $Re = 20$ and $Re = 39$, respectively. We underline that the action of the turbulent Reynolds stress is likely to be enhanced with increasing $Re$.

\begin{figure*}[ht!]
    \centering
    \includegraphics[width=\textwidth]{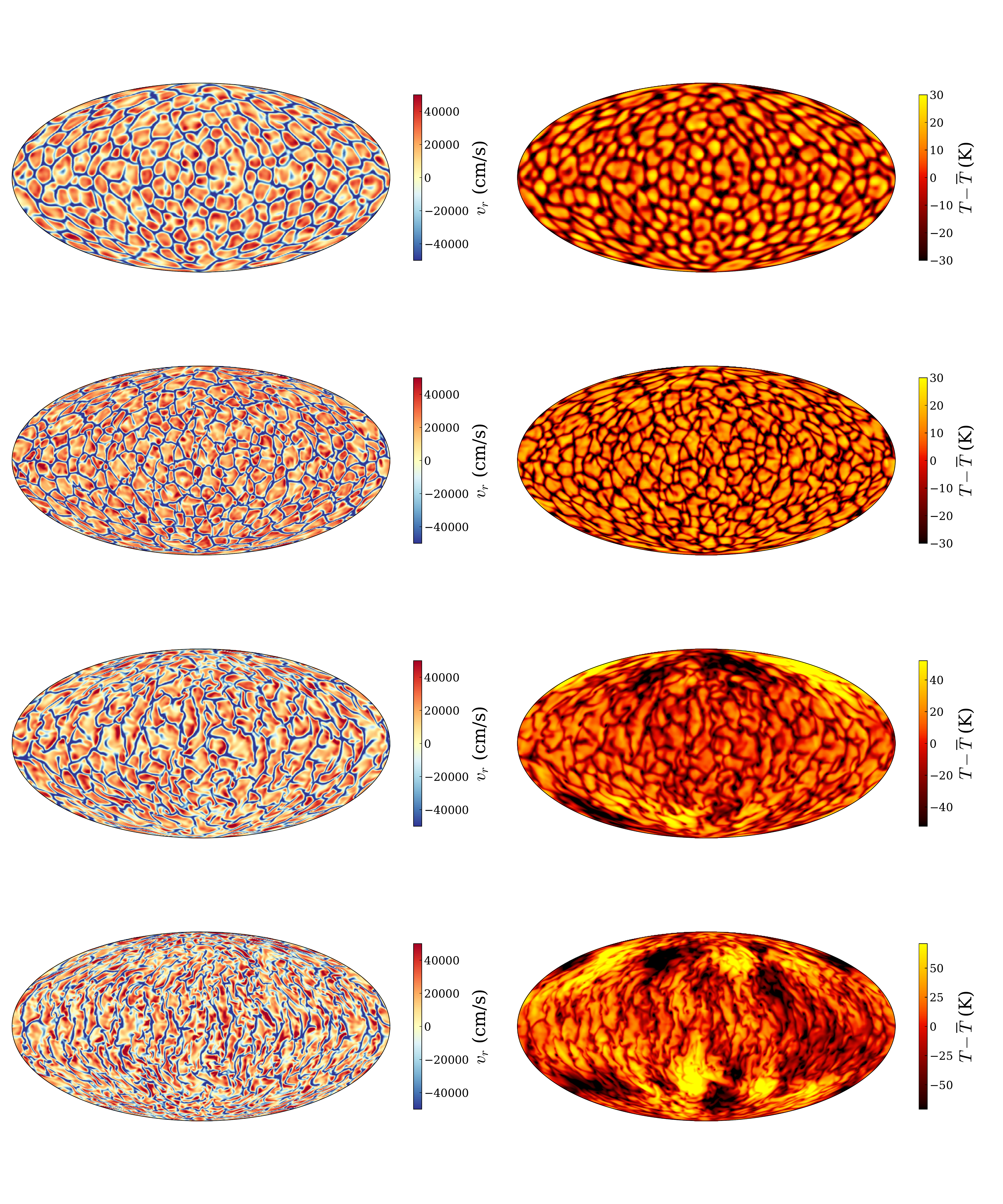}
    \caption{Radial velocity $v_r$ (left) and temperature perturbation $T-\overline{T}$ (\textit{right}) at depth $r = 0.95 \, R_\star$ for the F1a, F1b, F3, and F5 cases (top to bottom). The latitudinal mean temperature perturbation, $m=0$, is subtracted from $T$.
    }
    \label{fig:vr_T_shell_slices}
\end{figure*}

\begin{figure}[ht!]
    \centering
    \includegraphics[width=0.48\textwidth]{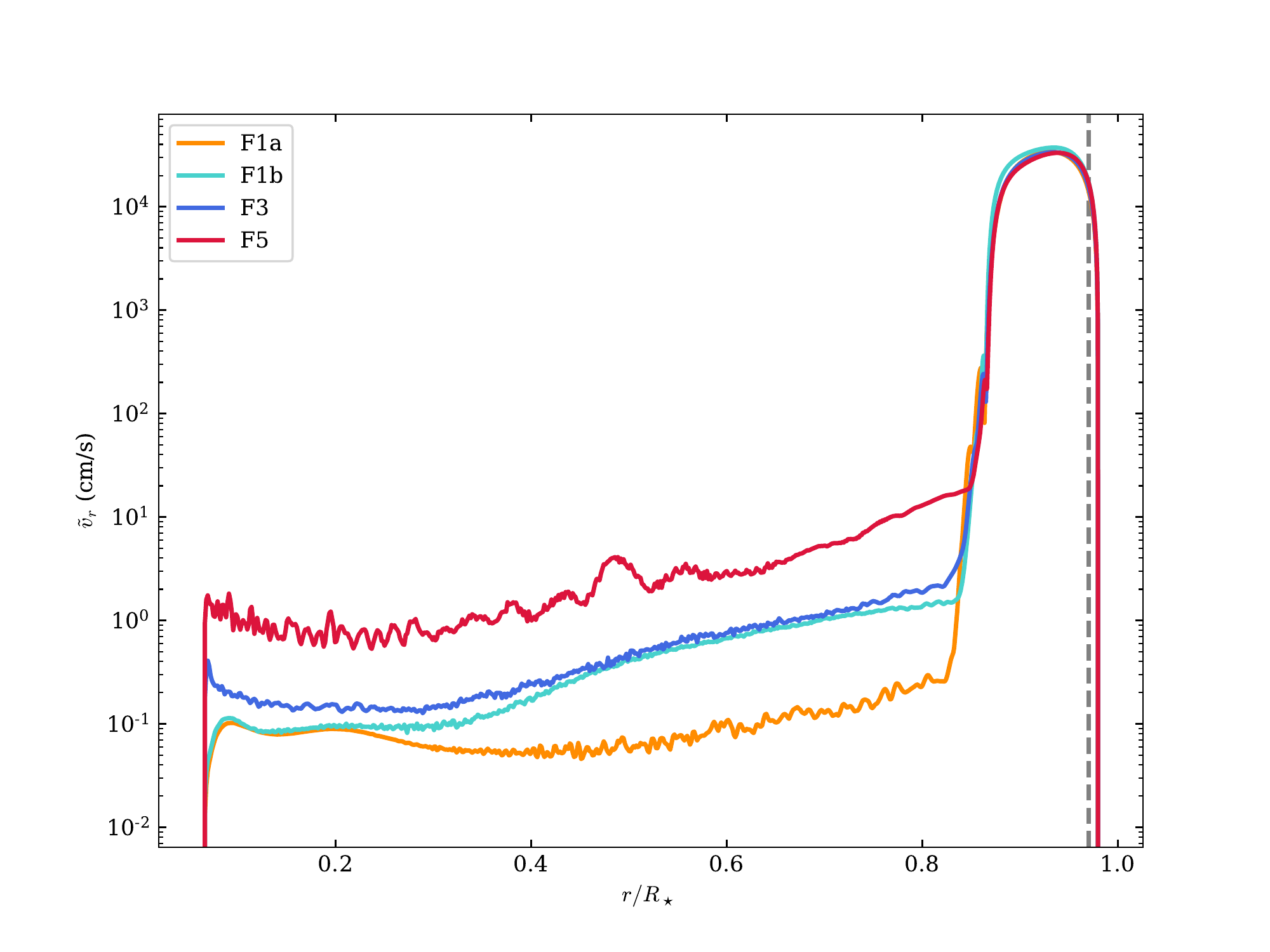}
    \caption{$\tilde{v}_r$ profiles for the F1a (orange), F1b (cyan), F3 (blue), and F5 (red) cases. The vertical dashed-grey line correspond to the depth where we examine the mode signatures for the F5 case in Sect.~\ref{sec:surface_visibility}.}
    \label{fig:vrms_shav}
\end{figure}

\begin{table}[h!]
    \centering
    \caption{Dimensionless numbers in the middle of the convective zone, $\tau_\mathrm{conv}$, and $r_c - r_0$ penetration depth. 
    }
    \begin{tabular}{ccccc}
    \hline \hline
    Case & F1a & F1b & F3 & F5 \\
    \hline
    $Re$ & 20 & 73 & 39 & 64 \\
    $Ro_f$ & 4.1 & 6.1 & 1.9 & 1.4 \\
    $Ro_s$ & 5.8 & 6.8 & 2.0 & 1.2  \\
    $Ro_c$ & 2.3 & 2.0 & 0.8 & 0.6 \\
    $Ra$ & \num{1.5e3} & \num{1.2e4} & \num{6.4e3} & \num{2.5e4} \\
    $Ta$ & \num{1.1e3} & \num{1.2e4} & \num{4.0e4} & \num{3.1e5} \\
    $Ra^*/Ra_c$ & 14.3 & 23.9 & 4.5 & 2.8 \\
    $\tau_\mathrm{conv}$ (day) & 4.8 & 4.1 & 4.8 & 4.8 \\
    $\overline{d}_\mathrm{ov}$ ($H_p$) & 0.22 & 0.23 & 0.18 & 0.15 \\
    \end{tabular}
    \label{tab:dimensionless_numbers}
\end{table}

Convection is only possible above a critical value of the Rayleigh number $Ra$ \citep[e.g.][]{Chandrasekhar1961,Jones2009} and, therefore, a sufficient level of supercriticality needs to be sustained in the simulations for convection to dominate diffusive phenomena. 
\citet{Takehiro2020} confirmed that, in the anelastic approximation the critical value of $Ra$ scaled as $Ta^{2/3}$, similarly to the Boussinesq case, where $Ta = 4 \Omega_0^2 L^4 / \nu^2$ is the Taylor number.
The diffusivity profiles, $\nu$ and $\kappa,$ should scale as $\Omega_0^{-2}$ in order to keep the same level of supercriticality in each simulation but the limitation of available computing resources prevents from actually adopting this scaling \citep[e.g.][]{Augustson2012}. 
Hence, we adopted, for the diffusivities, a scaling of $\Omega_0^{-0.63}$ between the F1a and F3 cases and $\Omega_0^{-0.75}$ between the F1a and the F5 cases, as a best compromise between numerical costs and physical constraints.
In order to estimate the supercriticality level of each of our models, we computed the modified Rayleigh number defined by \citet{Takehiro2020}, $Ra^*$, and we compared it with the critical value $Ra_c$ we obtain when taking their M11R5 model as reference for the scaling (for their model $Ta = \num{5.9e6}$ and $Ra_c = \num{4e5}$). We obtained $Ra^*/Ra_c = 14.3$, 23.9, 4.5, and 2.8 for the F1a, F1b, F3, and F5 cases, respectively, confirming that all of them are in a supercritical state. The supercriticality level achieved in the different cases is however significantly different, mainly because of the different rotation rates we impose. 
In order to assess how this influences the convection power spectrum, we therefore represent in Fig.~\ref{fig:convection_spectrum} the spherical harmonic decomposition of the time average of the rms radial velocity, $\tilde{v}_r$, for each case. The decomposition is summed over $m$ in the top panel and over $\ell$ in the bottom panel.
At low $\ell$, the F3 case exhibit the largest values, followed by the F5 and F1b cases.
It appears that the transition between the inertial range and the viscous-dominated domain happens in the $\ell = 20 - 50$ range. Beyond $\ell = 100$, the spectrum is completely dominated by viscous diffusion. The inhibiting effect of rotation can be distinguished around the peak at $\ell = 30$, where we note that the F1b decomposition peaks significantly higher than for the F5 case, although the two cases have identical $\kappa$ and $\nu$ profiles. As expected due to its fastest rotation rate, the F5 case peaks at higher $\ell$ \citep{Takehiro2020}.
Concerning the $m$ decomposition, the velocity spectrum is flat for for the F1a and F1b cases at low and intermediate $m$. 
It is flat only at low $m$ for the F3 and F5 cases, then increases and peaks between $m=10$ and $m=30$.
Around the $m \approx 30$ threshold, the velocity drastically decreases for all cases as $m$ increases. The $m$-spectrum reaches its maximal value for $m=17$, 8, 20, 25 for the F1a, F1b, F3, and F5 cases, respectively. We note that these maximal values increase with rotation, following the trend identified by \citet{Takehiro2020} for critical azimutal numbers.
We recall that due to the apparent mismatch with helioseismic solar convective velocity \citep{Hanasoge2012}, the so-called convective conundrum, absolute values for convective velocities obtained from simulations must be considered with care although the general trend they follow is consistent \citep{Hanasoge2016,Hotta2021,Brun2022}. The values of $Re$, $Ta$, $Ra^*/Ra_c$, and $\tau_\mathrm{conv}$ are summarised in Table~\ref{tab:dimensionless_numbers}.

Figure~\ref{fig:vr_T_shell_slices} shows the impact of rotation on the convective structures, represented through radial velocity, $v_r$, and temperature perturbation $T - \overline{T}$ maps at $r=0.95$ $R_\star$.
At this depth, $v_r$, and $T - \overline{T}$ are clearly correlated: downward flows carry negative temperature perturbations, while upward flows carry positive temperature perturbations. 
At 1 $\Omega_\odot$, the structures are only marginally affected by rotation, their shape does not depend much of the latitudinal position. Smaller scales are visible for the F1b case as it is more turbulent than F1a. 
On the contrary, when considering the F3 and F5 cases, the increase of $\Omega_0$ yields sharper patterns for the velocity field. Banana-shape cells akin to Busse's columns appear at this rotation rate \citep{Busse1970}, showing evidence that the convection dynamics is strongly affected by the rotation rate of the model and the strength of the Coriolis acceleration.
We notice that the structures tend to align perpendiculary to the equator. 
We note that in the F3 and F5 cases, large-scale structures with large temperature perturbations (positive or negative) appear at high latitude.
As it can be seen in Fig.~\ref{fig:vrms_shav}, the $\tilde{v}_r$ radial profile is similar in the convective zone for the four cases. 
In the middle of the convective zone, we have $\tilde{v}_r \approx \num{3.4e4}$, \num{3.7e4}, \num{3.3e4}, and \num{3.2e4} cm/s in the F1a, F1b, F3, and F5 cases, respectively. We therefore confirm that the most turbulent model is the one with the highest convective velocities.
The $\tilde{v}_r$ value in the radiative zone is associated with the amplitude of IGWs propagating in the stellar interior. Larger $\tilde{v}_r$ are reached in the radiative zone for the F3 and F5 cases. F3 and F5 have lower $\kappa$ value in the radiative zone. Therefore, waves are less damped by thermal dissipation. 
We also notice that, due to the deeper position of the transition radius for the diffusivity drop, the mean $\tilde{v}_r$ stay at a level equivalent to the F3 case until $r/R_\star \approx 0.6$. At the bottom of the radiative zone, however, $\tilde{v}_r$ values from the F3 case are comparable to the F1b case.  

Finally, it should be reminded that convection is intriscally a dynamical process, with cells evolving over time. The interaction and combination between turbulent flows shape upwards and downwards travelling convective structure. Downwards flows with the largest amount of power give rise to the convective plumes interacting with the radiative zone. The plumes act similarly to a piston as they overshoot in the radiative and inject power in the inner regions in the form of IGWs.   

\subsection{Differential rotation in the convective zone and the sharp tachocline \label{sec:rot_diff}}

We now turn to considering the differential rotation regimes achieved in the simulations.
Following \citet{2017ApJ...836..192B}, we define the fluid Rossby number as: 
\begin{equation}
    Ro_f = \frac{|\nabla \times \bm{v}|}{2\Omega_0} \; ,
\end{equation}
where $\nabla \times \bm{v}$ is the vorticity of the flow. \citet{2017ApJ...836..192B} expect a transition between the solar (fast equator, slow poles) and anti-solar rotation (slow equator, fast poles) regime at $Ro_f \approx 1$ \citep[see also][]{Gastine2014,Guerrero2019,Warnecke2020}. At very low $Ro_{f}$, as a consequence of the Taylor-Proudman theorem, the rotational regime becomes cylindrical. 

We also compute the stellar Rossby number $Ro_s$ \citep[e.g.][]{Noyes1984,Corsaro2021} and the convective Rossby number $Ro_c$ \citep{Gilman1981}. Here, $Ro_s$ is given by
\begin{equation}
   Ro_s = \frac{P_\mathrm{rot}}{\tau_\mathrm{conv}} \; ,
\end{equation}
where the rotation period is $P_\mathrm{rot} = 2 \pi / \Omega_0$, and $\tau_\mathrm{conv}$ is the convective turnover time that we compute as the ratio of the thickness of the convective zone in the simulation domain, $d_\mathrm{CZ}$, and the mean $\tilde{v}_r$ in the convective zone, $\langle \tilde{v}_r \rangle_\mathrm{CZ}$.
For $Ro_c$, we use the following definition
\begin{equation}
   Ro_c = \sqrt{\frac{Ra}{Ta Pr}} \; , 
\end{equation}
where $Ra = (- \partial \rho / \partial S) (\partial S_\mathrm{tot} / \partial r) g L^4 / \rho \kappa \nu$, with $S_\mathrm{tot} = \bar{S} + S$. Here, again, for the characteristic length, $L$, we consider $d_\mathrm{CZ}$. As expected by \citet{2017ApJ...836..192B}, $Ro_f$, $Ro_s$, and $Ro_c$ scale in a similar way, except for the $Ro_c$ value of the F1b case, which is smaller than for the F1a case. These values are  between 3 and 6 for the 1 $\Omega_\odot$ cases and approach unity for the fastest rotating cases. 

The values of $Ro_f$, $Ro_s$, $Ro_c$, and $Ra$ in the middle of the convective zone are summarised for each case in Table~\ref{tab:dimensionless_numbers}. 

\begin{figure}[ht!]
    \centering
    \includegraphics[width=0.48\textwidth]{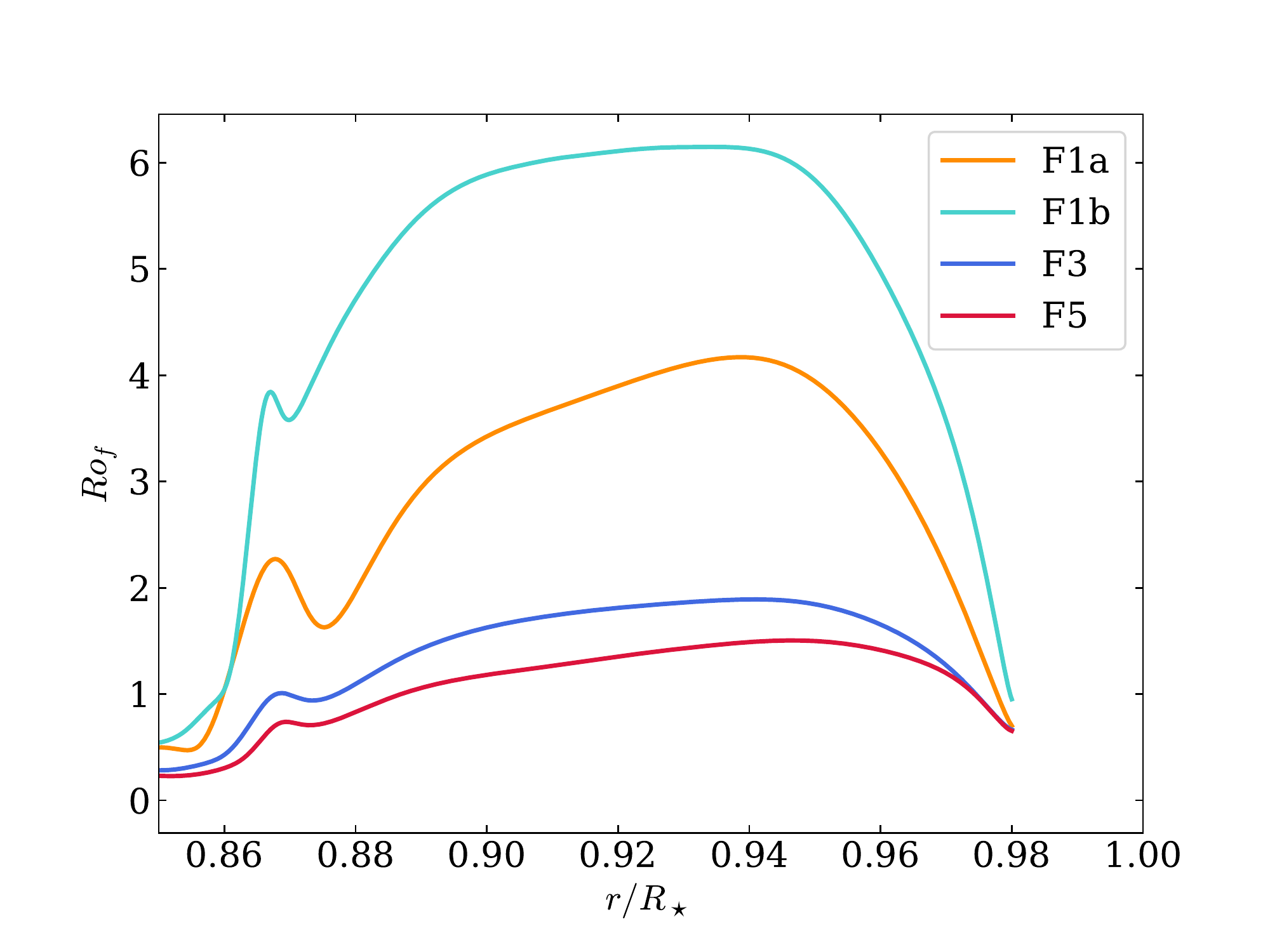}
    \caption{Fluid Rossby number $Ro_{f}$ profiles in the convective zone for the F1a (orange), F1b (cyan), F3 (blue), and F5 (red) cases.}
    \label{fig:rossby}
\end{figure}

\begin{figure}[ht!]
    \centering
    \includegraphics[width=0.48\textwidth]{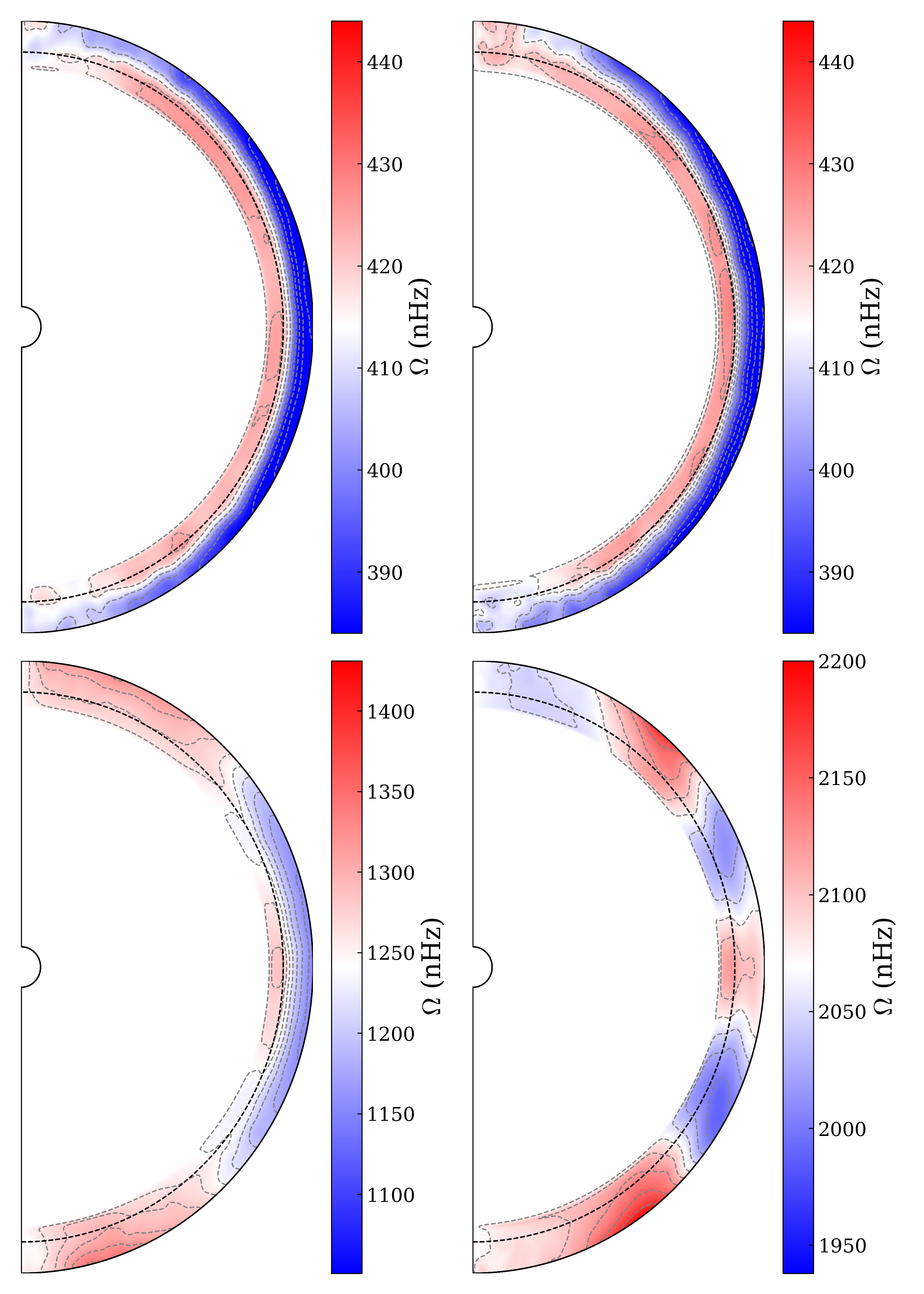}
    \caption{Differential rotation for F1a (\textit{top left}), F1b (\textit{top left}), F3 (\textit{bottom left}), and F5 (\textit{bottom right}) cases. Retrograde and prograde (relatively to the co-rotational frame) flows are shown in blue and red, respectively. The dashed black line corresponds to $r_\mathrm{CZ}$ and the dashed grey lines are isocontours..
    }
    \label{fig:rotdiff}
\end{figure}

We present the $Ro_{f}$ profile in the convective zone in Fig.~\ref{fig:rossby}.
Considering the $v_\phi$ component of the velocity field averaged over longitude and time, we also computed the differential rotation state for the different cases. The $(r, \theta)$ differential rotation profiles are represented in Fig.~\ref{fig:rotdiff}.  
For the F1a and F1b cases, the simulation exhibits a shellular rotation profile which is consistent with the fact that $Ro_{f}$ in the middle of the convective zone is significantly larger than 1, as seen in Fig.~\ref{fig:rossby}. The F3 case behaviour is still clearly anti-solar, with a $Ro_{f}$ of 1.9 in the middle of the convective zone.
Interestingly, the transition from prograde to retrograde flows (in the co-rotational frame), at every latitude for F1a and F1b and close to the equator for F3, intervene close to $r_\mathrm{CZ}$.
The F5 case has R$_{of} \approx 1.4$ in the middle of the convective zone and the structure of the flows observed in Fig.~\ref{fig:rotdiff} suggest that the model is in a transitional regime from an anti-solar to a solar differential rotation regime.    
In our F5 case, we notice an asymmetry between behaviours at high latitudes, with slow flows close to the north pole and fast flows close to the south pole. Faster rotating F-star models published by \citet{Augustson2012} confirm the change of rotation regime towards Taylor-Proudman constrained states for low $Ro$. We recall that their 1.3~$M_\odot$ model rotating at 10~$\Omega_\odot$ exhibits a $Ro_f$ value of 0.84 and a solar differential rotation regime, which is consistent with the predictions from \citet{2017ApJ...836..192B}. 

\subsection{Overshoot}

\begin{figure*}[ht!]
    \centering
    \includegraphics[width=\textwidth]{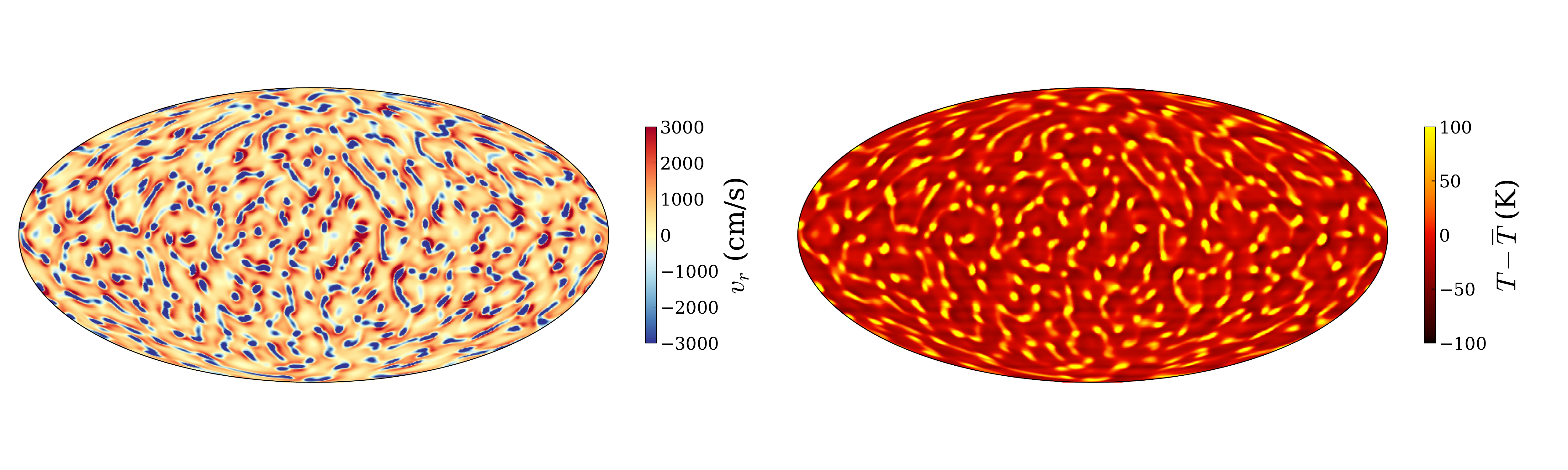}
    \caption{Radial velocity $v_r$ (\textit{left}) and temperature perturbation $T-\overline{T}$ (\textit{right}) at depth $r = 0.86 \, R_\star$ for the F1a case. The latitudinal mean temperature perturbation was substracted from $T$.
    }
    \label{fig:shell_slices_vr_T_F1a_depth_11}
\end{figure*}

The convective motions overshoot into the radiative interior, leading to mixing and IGWs generation.
In Fig.~\ref{fig:shell_slices_vr_T_F1a_depth_11}, we illustrate how the correlation between $v_r$ and $T - \overline{T}$ change in the overshoot region relatively to the convective zone. We show the $v_r$ and $T - \overline{T}$ maps at depth $r = 0.86 \, R_\star$ for the F1a case. 
In the convective zone, $v_r$ and $T - \overline{T}$ were correlated, that is, the upwards flows were related to positive temperature perturbations and reciprocally. On the contrary, in the overshoot region, the two quantities are now anti-correlated: downwards travelling plumes penetrating the top of the stably-stratified radiative zone are associated with positive temperature perturbations. This is expected from our understanding of penetrative convection in stellar interiors \citep{1991A&A...252..179Z,Brummell2002}.

\begin{figure}[ht!]
    \centering
    \includegraphics[width=.49\textwidth]{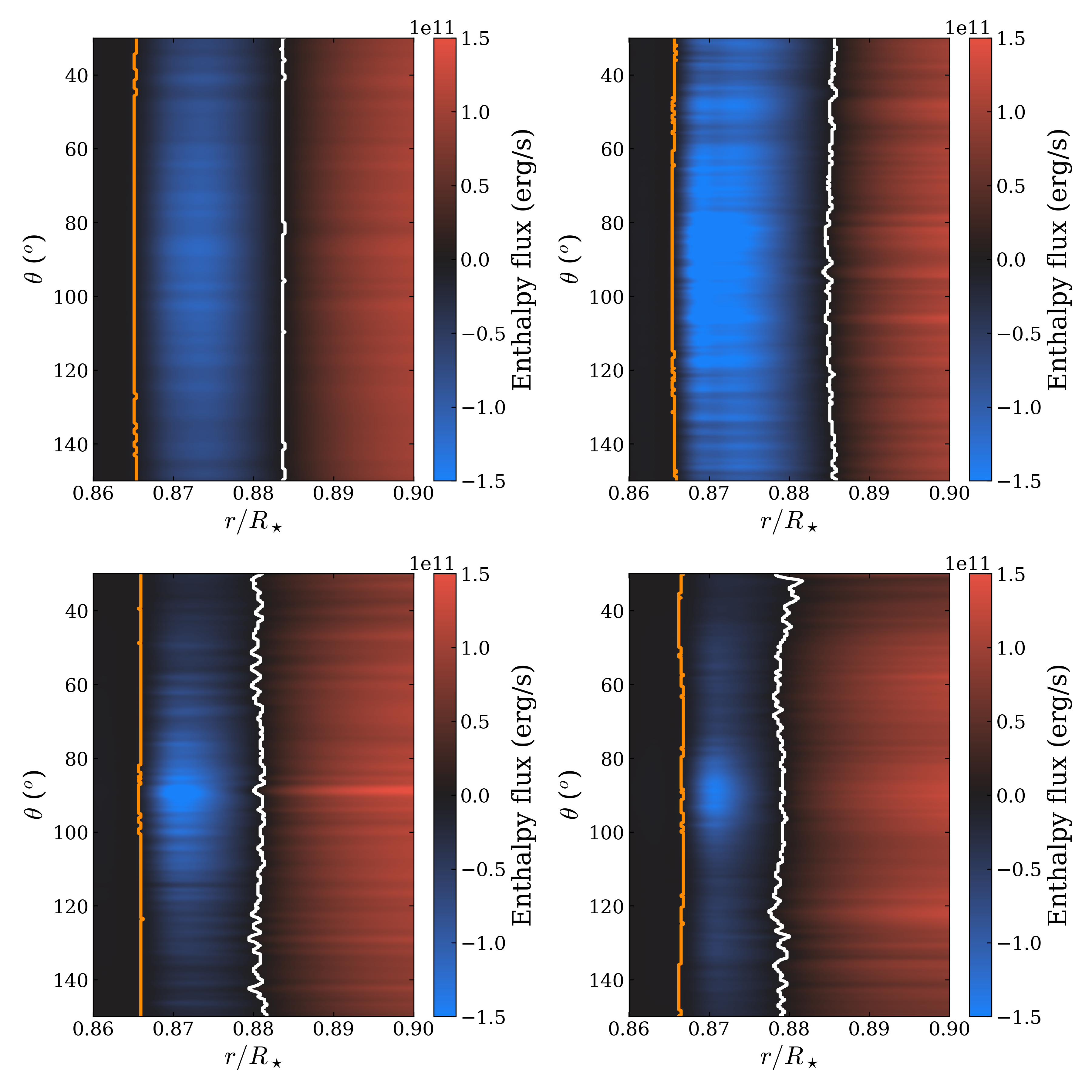}
    \caption{Overshoot region (\textit{blue}) and bottom of the convective zone (\textit{red}) in the F1a (\textit{top left}), F1b (\textit{top right}), F3 (\textit{bottom left}), and F5 (\textit{bottom right}) models, for $\theta$ spanning from 30 to 150 $^o$. The white line show the radius $r_c$ at which the enthalpy flux crosses zero at a given latitude while the orange line signals the radius $r_0$ where the local enthalpy flux is equal to a tenth of the maximal enthalpy flux at the corresponding latitude.}
    \label{fig:overshoot}
\end{figure}

Figure~\ref{fig:overshoot} shows the enthalpy flux at the bottom of the convective zone and in the overshoot region for the F1a, F1b, F3, and F5 cases. Following \citet{2011ApJ...742...79B,2017ApJ...836..192B}, we computed, for every latitude, the boundaries $r_c$ and $r_0$ where the enthalpy flux becomes negative and where the enthalpy flux is only 10~\% of the maximal absolute value in the overshoot region at this latitude, respectively. We note that the shape of the $r_0$ profile does not change significantly with the rotation, however, the $r_c$ profile differs significantly in the F3 and F5 cases compared to the F1b case. We also note for fast rotating models, the enthalpy flux intensity in the overshoot region is concentrated towards the equator likely due to the stronger alignment of convective rolls for these stars. By computing the difference of the mean value of $r_c$ and $r_0$ at latitudes below 55$^o$ for each case, we obtain the mean penetration depth of the overshooting flows, $\overline{d}_\mathrm{ov}$, in the same way as \citet{2017ApJ...836..192B}. We obtain $\overline{d}_\mathrm{ov} = 0.22$, 0.23, 0.19 and 0.15 $H_p$ where $H_p \approx \num{8.8e9}$~cm is the pressure scale height at the base of the convective zone. 
The comparison between F1a and F1b shows that, at constant $\Omega_0$, the overshoot depth increases with $Re$ but, as expected, the penetration depth of the overshooting flow is significantly reduced as rotation increases.
The $\overline{d}_\mathrm{ov}$ values are summarised together with the dimensionless numbers in Table~\ref{tab:dimensionless_numbers}.
These overshooting motions contribute to the excitation of the IGWs. In particular, \citep{Pratt2017} showed that two characteristic layers of penetration could be distinguished for overshooting plumes, the deepest one corresponding to the excitation region of the IGWs.


\section{IGW properties \label{sec:igw}}

In this section, we study in detail the properties of the IGWs that are generated by the interaction of the convective motions with the top of the radiative zone. The dispersion relation for IGWs is \citep{Press1981}: 
\begin{equation}
\label{eq:dispersion_relation_no_rot}
    \omega^2 = N^2 \frac{k_h^2}{k^2} \; ,
\end{equation}
where $k$ is the norm of the wave vector $\bm{k} = \bm{k}_r + \bm{k}_h$. The norm of the horizontal wave vector $\bm{k}_h = \bm{k}_\theta + \bm{k}_\phi$ is: 
\begin{equation}
    k_h = \frac{\sqrt{\ell (\ell+1)}}{r}
.\end{equation}
Concerning gravito-inertial waves, when the Coriolis parameter $f = 2\Omega_0$ is such that $f \ll N$, the effect of the Coriolis acceleration on the wave behaviour is negligible and we retrieve the pure-gravity case. Hence, we expect high-frequency waves propagation not to be affected by the rotation regime of the different cases. However, as we increase $\Omega_0$, the range of waves significantly affected by inertial effects expands.

\subsection{Raytracing}

\begin{figure}[ht!]
    \centering
    \includegraphics[height=.8\textheight]{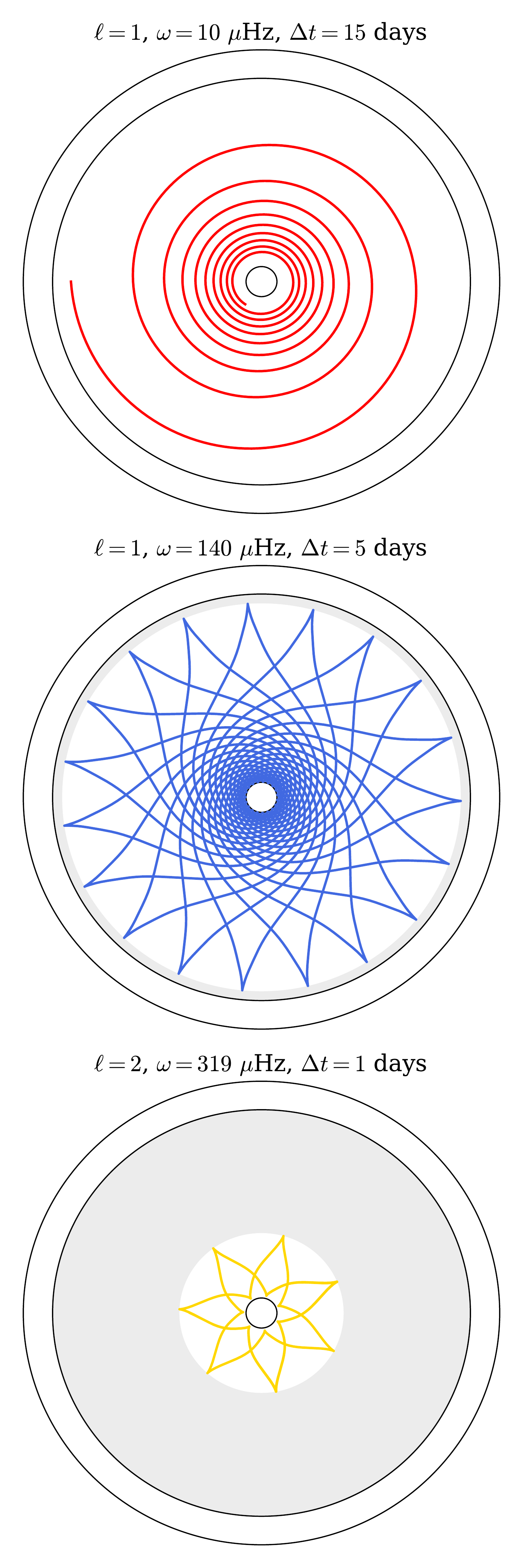}
    \caption{Propagation path computed with ray-tracing for three IGWs: $\ell=1$, $\omega = 10$ $\mu$Hz (\textit{top}), $\ell=1$, $\omega = 140$ $\mu$Hz (\textit{middle}), $\ell=2$, $\omega = 319$ $\mu$Hz (\textit{bottom}). The integration time, $\Delta t,$ considered for each ray is specified.
    From the surface to the center, the black circles represent the position of the stellar surface, the bottom of the convective envelope and the edge of the convective core. The grey area corresponds to the part of the radiative zone where the wave is evanescent.}
    \label{fig:raytracing}
\end{figure}

In order to better understand the expected IGWs behaviour at different frequencies for our F-type model, we start by computing the propagation path of IGWs at different frequencies following a ray-tracing Hamiltonian method \citep{Gough1993}. As $k_h$ depends only on $\ell$ and $r$, we set $\theta = 0$ and $k_\theta = 0$ and we consider only a 2D problem in the equatorial plane. However, as $\theta$ and $\phi$ can be freely interchanged in this specific case, we could have chosen to place ourselves in any meridian plane without any modification to our subsequent analysis. To perform the ray-tracing, we must numerically solve the following set of equations:
\begin{equation}
\label{eq:hamiltonian_raytracing}
    \left \{
    \begin{aligned}
     \diff{r}{t} &= \diffp{W}{r} \; , \\
     \diff{\phi}{t} &= \frac{1}{r}\diffp{W}{\phi} \; , \\
     \diff{k_r}{t} &= - \diffp{W}{{k_r}} + \frac{1}{r} \diffp{W}{{k_\phi}} k_\phi \; , \\
     \diff{k_\phi}{t} &= - \frac{1}{r} \diffp{W}{{k_\phi}} - \frac{1}{r} \diffp{W}{{k_r}} k_\phi \; , \\
    \end{aligned}
    \right . 
\end{equation}
where $W (r, \theta, k_r, k_\phi) = \omega$ is the Hamiltonian of the considered system. Using the dispersion relation of Eq.~\ref{eq:dispersion_relation_no_rot}, we compute the propagation path for three IGWs with different properties, $\ell=1$ at $\omega = 10$ $\mu$Hz,  $\ell=1$ at $\omega = 140$ $\mu$Hz,  $\ell=2$ at $\omega = 319$ $\mu$Hz. The integration times, $\Delta t$, are (respectively) 15, 5, and 1 days. The radial resonant cavities corresponding to the three chosen frequencies are represented along with the $N$ profile in Fig.~\ref{fig:bv}.
The obtained propagation paths are shown in Fig.~\ref{fig:raytracing}. 
As expected from the $N$ profile shown in Fig.~\ref{fig:bv}, high-frequency IGWs are trapped in the deepest regions of the radiative zone. High-frequency IGWs propagate much faster in the radiative interior. Low-frequency IGWs have a characteristic spiralling trajectory and take more time to reach the edge of the convective core.     


\begin{figure}[ht!]
    \centering
    \includegraphics[width=.49\textwidth]{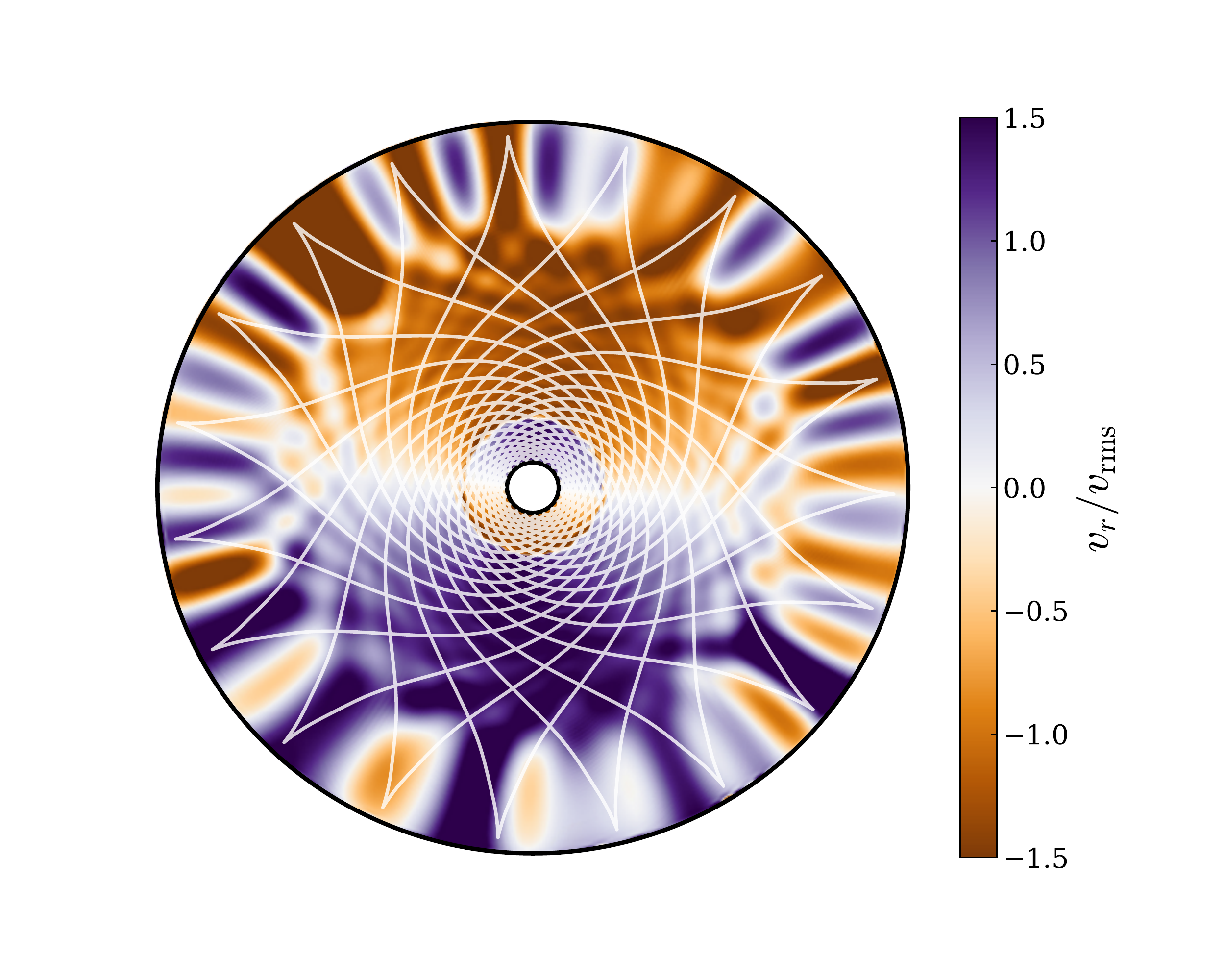}
    \caption{Equatorial cut for $v_r/\tilde{v}_r$ for the radiative zone in the F1a case, with application of a temporal filtering centered at 140 $\mu$Hz. The 2D propagation ray obtained by numerically solving the system of Eq.~\ref{eq:hamiltonian_raytracing} and shown in the middle panel of Fig.~\ref{fig:raytracing} is overplotted in white. We note that there is a good qualitative agreement.}
    \label{fig:filtering_140_muHz}
\end{figure}

In order to present more evidence of IGWs propagating in the radiative zone of our simulations, we considered the $v_r$ temporal evolution in an equatorial cut of the F1a case. We applied a finite impulse response (FIR) filter with a passband centered at 140 $\mu$Hz in order to isolate the $n=2$, $\ell=1$ mode. We compare in Fig.~\ref{fig:filtering_140_muHz} the result of the filtering with the propagation path obtained by solving the Hamiltonian system. Considering the sign of $v_r / \tilde{v}_r$, the dipolar structure of the mode oscillation is clearly visible after the filtering, as well as the position of the two radial nodes of the mode. What we see in this filtering can be interpreted as the pattern of a ray interfering with itself \citep{Gough1993}. Due to the diffusive effects related to the $\kappa$ and $\nu$ profiles, the characteristic shape of the mode spreads when compared to the path obtained with the ray-tracing. However, having such a good qualitative agreement between our filtered model and the shape of the waves obtained with the raypath theory gives us confidence in the realism of internal waves excited in our simulations.   

\subsection{IGWs power spectrum \label{sec:igw_spectrum}}

To obtain the IGW power spectrum, we produced outputs of $v_r (r, \theta, \phi, t)$ at a mean sampling d$t$ low enough to have $N_\mathrm{max} < \omega_N$, with $\omega_N$ the Nyquist frequency of the output signal and $N_\mathrm{max}$ the maximal value of the $N$ profile. In our case, this means that we are careful to have d$t \leq 1250$~s. The length of the time series we use for each case are given in Table~\ref{tab:model_summary}.
We expand $v_r (r, \theta, \phi, t)$ into a spherical harmonics representation to obtain $v_r(r, \ell, m, t)$. This representation is more suitable to study the modes as their properties directly depend of $\ell$ and $m$ (as for example mode period spacings or rotation splittings). From the Fourier transform $\tilde {v}_r (r, \ell, m, \omega)$ of $v_r$, we then obtain the power spectrum $E_{\ell,m}$ as
\begin{equation}
\label{eq:ref_Elm}
    E_{\ell,m} (r, \omega) = |\tilde{v_r} (r, \ell, m, \omega)|^2 \; ,
\end{equation}
and, by summing over the $m$ for each $\ell$, we have the power spectrum, $E_{\ell}$, 
\begin{equation}
    E_{\ell} (r, \omega) = \sum_{m=-\ell}^{\ell} E_{\ell,m} (r, \omega) \; .
\end{equation}
Some additional practical details concerning the way we obtain $E_{\ell,m}$ and $E_\ell$ are provided in Appendix~\ref{appendix:sph_expansion}.

\begin{figure*}[ht]
    \centering
    \includegraphics[width=\textwidth]{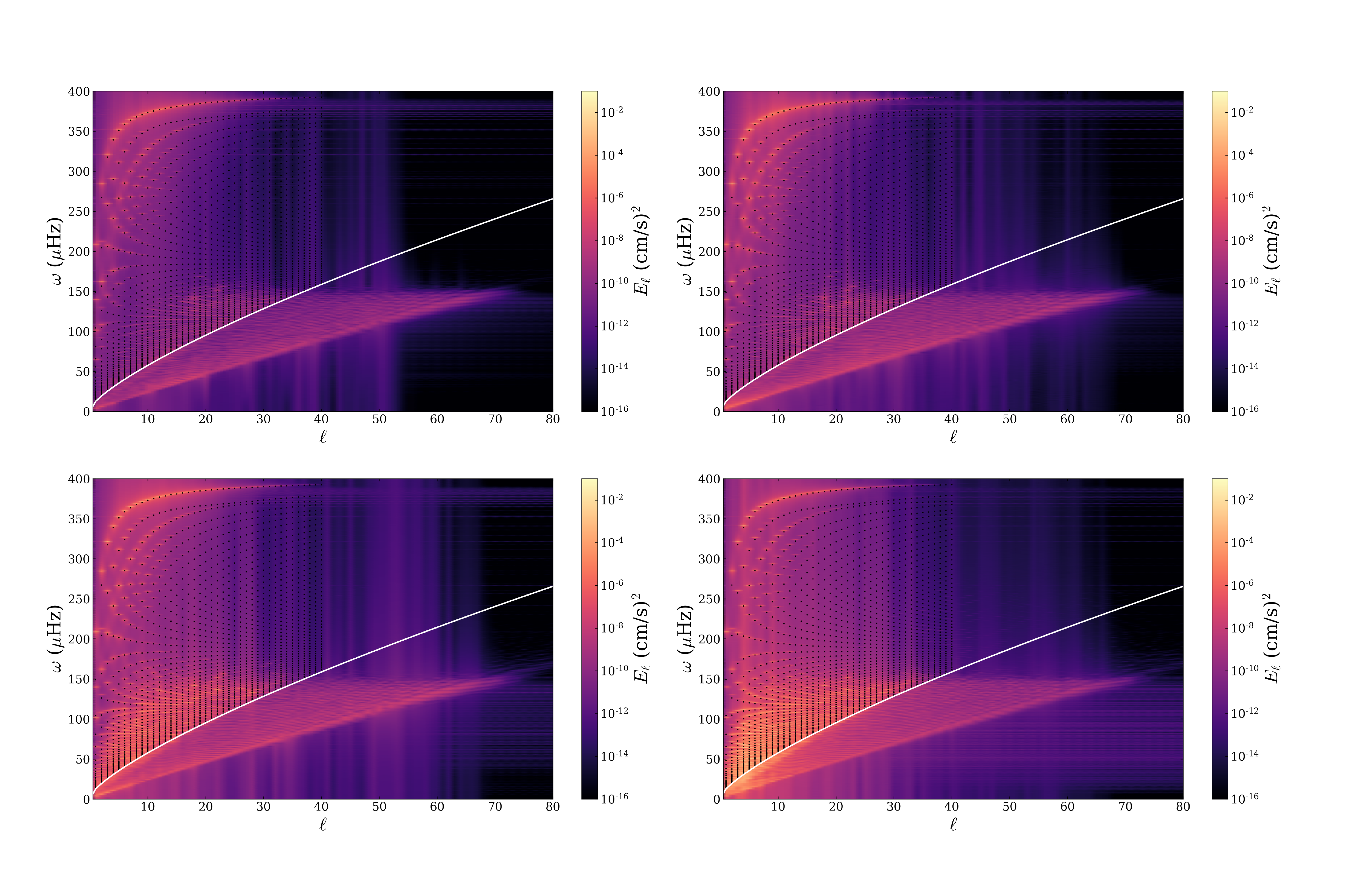}
    \caption{$E_\ell$ power spectrum at $r=0.18 \, R_\star$ for the F1a (\textit{top left}), F1b (\textit{top right}), F3 (\textit{bottom left}),  and F5 (\textit{bottom}) case. The mode frequencies computed with GYRE are overplotted (black dots) for $\ell=1$ to $\ell =40$. In each panel, the white line highlights the separation between the g-mode region (\textit{top left}) and the progressive waves region (\textit{bottom right}).}
    \label{fig:E_l_depth_80}
\end{figure*}




The $E_\ell$ power spectra at $r=0.18 \, R_\star$ obtained for the F1a, F1b, F3, and F5 cases are represented in Fig.~\ref{fig:E_l_depth_80}. 
The IGWs spatial damping rate in a non-adiabatic medium increases with $\ell$ and decreases with $\omega$ \citep[e.g.][]{1997A&A...322..320Z}. 
As pointed out by \citet{2015A&A...581A.112A}, two regions can therefore be distinguished in each spectrum: the bottom right part of the spectrum (high degree and/or low frequency) shows progressive waves which are significantly damped before reaching the edge of the convective core while in the upper left part of the spectrum (low degree or high frequency, or both), IGWs reflect on the core edge with enough amplitude to form standing waves, or g modes. In this region, g-modes are distributed along ridges in relation to their radial order, $n$. Following \citet{Ahuir2021}, we characterise the frequency cutoff between the two regions, $\omega_c$, through the following relation 
\begin{equation}
    \omega_c (\ell) = [\ell (\ell + 1)]^{3/8} \left[\int_{r_\mathrm{bottom}}^{r_\mathrm{CZ}} \kappa \frac{N^3}{\tau_c r^3} dr \right]^{1/4} \; ,
\end{equation}
where we find that adopting a critical damping parameter $\tau_c = 0.02$ correctly describe the cutoff profile observed in Fig.~\ref{fig:E_l_depth_80}.

As expected from Fig.~\ref{fig:vrms_shav} and the various level of $\tilde{v}_r$ in the radiative interior, the IGWs mean amplitude is significantly larger in F5 and F3 than F1a. In particular, low-frequency-mode amplitudes are significantly larger in the F3 and F5 cases.
Being a more turbulent version of F1a, F1b exhibits high-frequency modes of larger amplitude than in any other case, but the excitation of low-frequency modes is similar to what is observed in F1a. 

\begin{figure}[ht!]
    \centering
    \includegraphics[width=0.49\textwidth]{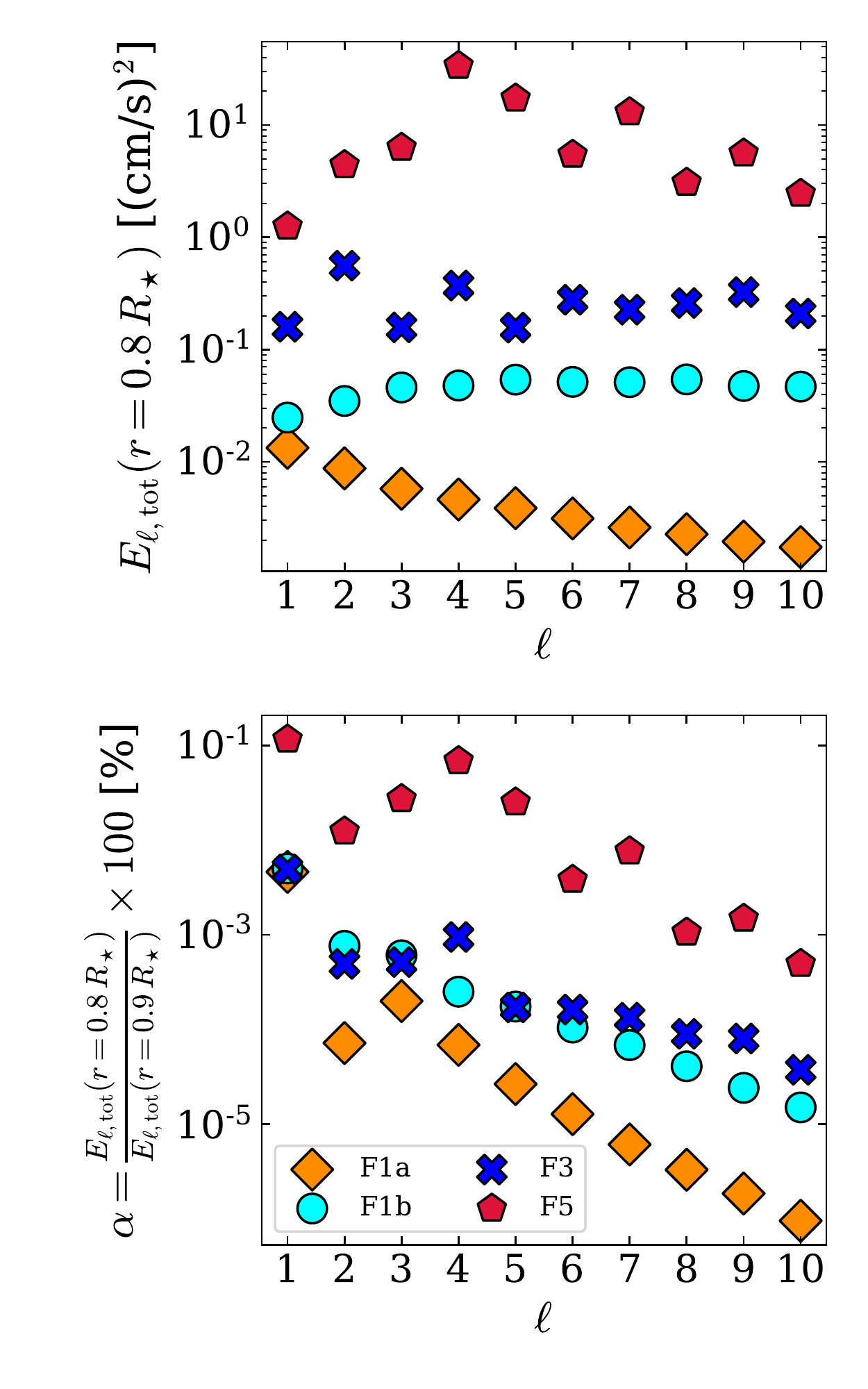}
    \caption{Amount of power injected in IGW. \textit{Top:} Total power $E_{\ell, \mathrm{tot}}$ for $1 \leq \ell \leq 10$ at $r=0.8$ $R_\star$ for the F1a (orange diamonds), F1b (cyan circles) F3 (blue crosses), and F5 (red pentagons) cases. \textit{Bottom:} Ratio between $E_{\ell, \mathrm{tot}} (r=0.8 \, R_\star),$ the radiative zone, and $E_{\ell, \mathrm{tot}} (r=0.9 \, R_\star),$ convective zone. The same symbols and colour coding are used as in previous figures.}
    \label{fig:power injection}
\end{figure}

In order to compare the mode excitation rate degree by degree, we compute the power index $E_{\ell, tot}$, taken as the summation of the $E_\ell$ component over the frequency bins, $\omega_i$,
\begin{equation}
    E_{\ell, \mathrm{tot}} (r) = \sum_{\omega_i} E_\ell (r, \omega_i) \; .
\end{equation}
For $1 \leq \ell \leq 10$, we represent in Fig.~\ref{fig:power injection} the values obtained for $E_{\ell, \mathrm{tot}}$ at depth $r=0.8 \, R_\star$, in the top of the radiative zone, as well as the ratio $\alpha = 100 \times E_{\ell, \mathrm{tot}} (r=0.8 \, R_\star) / E_{\ell, \mathrm{tot}} (r=0.9 \, R_\star)$. This allows us to compare on one hand the relative excitation level of each degree $\ell$, and to evaluate on the other hand the efficiency of power injection from the convective motions to the IGWs. As expected from it being more turbulent than F1a, the excitation level of the waves is more important in F1b than in F1a. The excitation level decreases with $\ell$ in F1a while it is relatively flat in $\ell$ for F1b, suggesting that intermediate degree $\ell$ have been more efficiently excited by the more turbulent flows of F1b (see Fig.~\ref{fig:convection_spectrum}). 
As already shown by Fig.~\ref{fig:E_l_depth_80}, the excitation level of the IGWs is the highest in F5, with a peak at $\ell = 4$. 
It is remarkable to note that despite the F3 case being more dissipative than the F1b case, the transmission of power and the wave excitation is more important.
It is interesting to note that, for the F5 case, the $\alpha$ ratio is the highest for $\ell = 1$ (with $\alpha = 0.12 \, \%$), suggesting an efficient transfer from convection, but remains the degree with the lowest excitation level in this case. For each case, $\alpha$ tends to decrease as $\ell$ increases. In the F5 case, the power injection peaks at $\ell = 4$. The enhanced power injection in the IGWs as $Ro_c$ (see Table~\ref{tab:dimensionless_numbers}) decreases has been predicted by \citet{2020ApJ...903...90A} and the result we present in Figs.~\ref{fig:E_l_depth_80} and \ref{fig:power injection} are in agreement with their theoretical considerations. 
However, the fact that intermediate $\ell$ are more excited than low $\ell$ might be considered surprising as modes of higher $\ell$ have increased inertia and require more energy from convection to be excited \citep{Provost2000}. 
As Fig.~\ref{fig:convection_spectrum} shows, the intermediate $\ell$ have increased convective velocities compared to low $\ell$. The absolute power in convection is nevertheless not a sufficient explanation alone as the slowly rotating cases have a similar  convective spectrum in the low to intermediate $\ell$. The mechanism enhancing power injection from rotation into the modes seems particularly efficient for intermediate $\ell$.

We will discuss in more detail the surface amplitudes and signatures of the modes in Sect.~\ref{sec:surface_visibility}.

\subsection{g-modes frequencies and eigenfunctions comparison with outputs from a 1D oscillation code \label{sec:gyre_comparison}}

We use the GYRE code \citep{2013MNRAS.435.3406T,2018MNRAS.475..879T,2020ApJ...899..116G} to compute the expected oscillation frequencies from the 1D input ASH profiles, for $\ell=1$ to $\ell=40$, in a case without rotation (referred later as the $\Omega_\star = 0$ GYRE run). 
As shown by Fig.~\ref{fig:E_l_depth_80}, the GYRE computed frequencies are globally in good agreement with what we observe in the 3D simulations. In particular, we can clearly see that the ridge structure for modes of different degrees $\ell$ but same orders $n$ coincide in the 3D simulation and in the GYRE computation.   

\begin{figure}[ht!]
    \centering
    \includegraphics[width=0.49\textwidth]{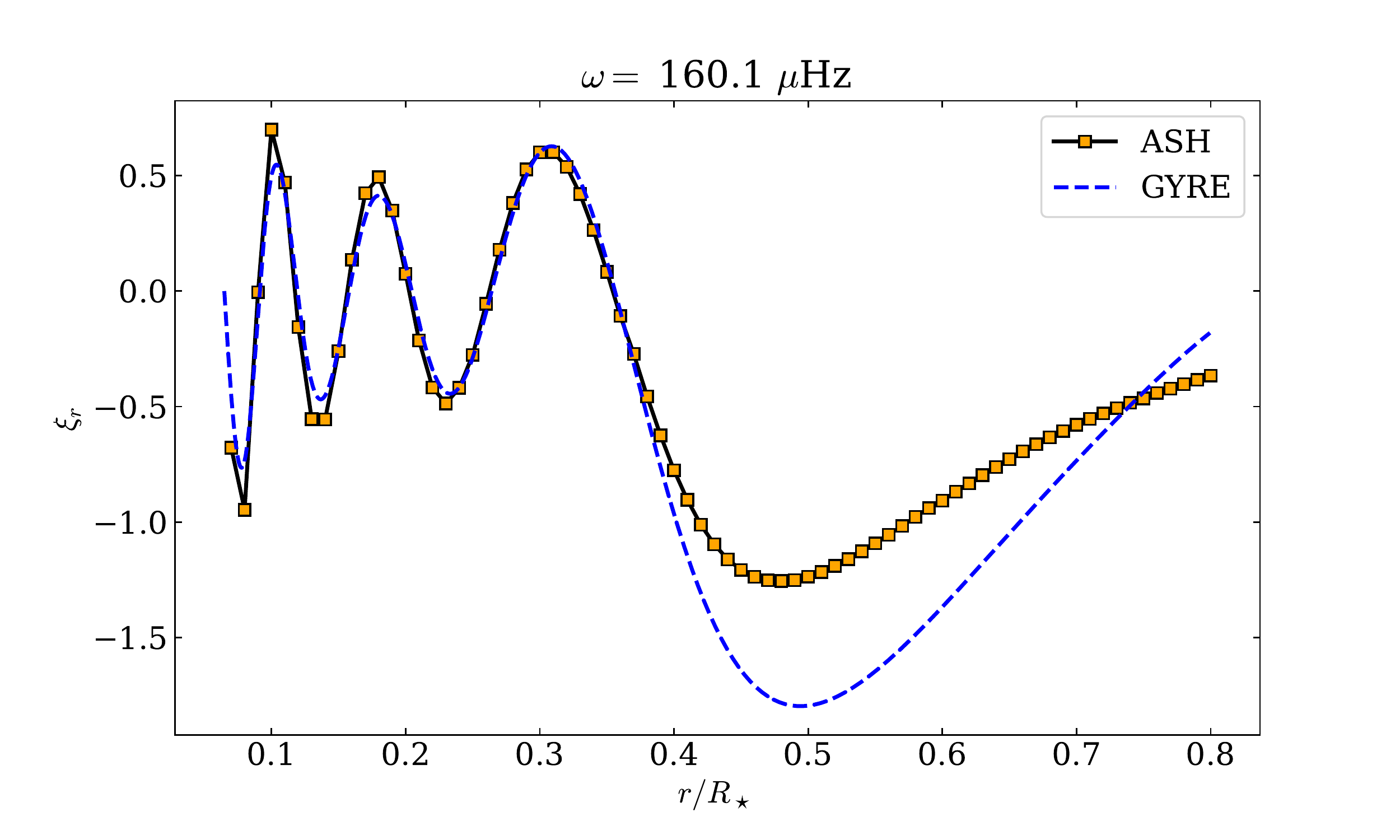}
    \includegraphics[width=0.49\textwidth]{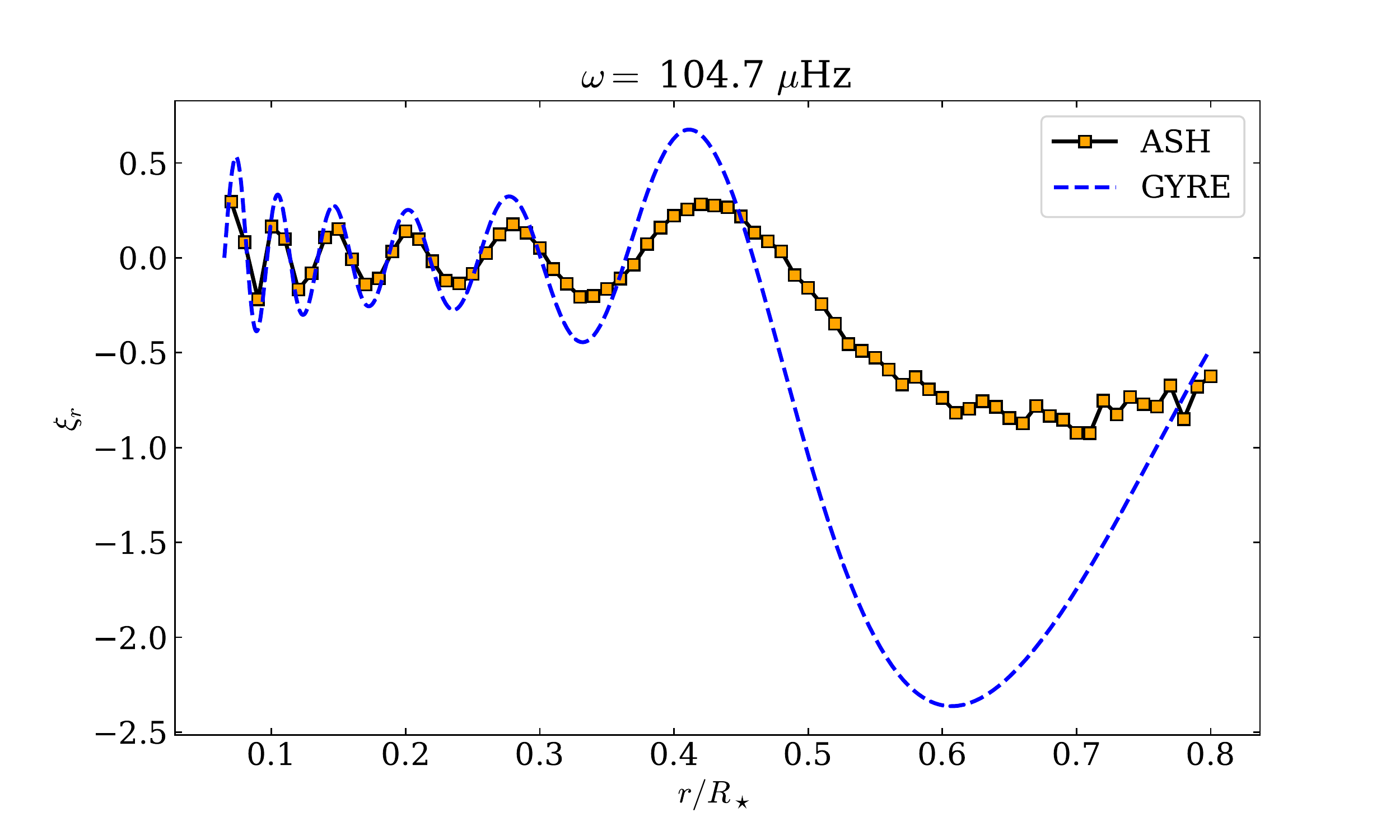}
    \caption{GYRE $\xi_r$ eigenfunctions (\textit{blue}) compared with ASH outputs for $\ell =5$, $n=7$ and $\ell =5$, $n=12$. The depths where ASH outputs are sampled are signaled by orange squares.}
    \label{fig:eigenfunctions}
\end{figure}

We also compare for some modes the $\xi_r$ displacement eigenfunctions with the outcomes of the 3D simulations. The eigenfunctions for $\ell=5$, $n=7$ and $n=12$ are shown in Fig.~\ref{fig:eigenfunctions}. The node position for these modes is in good agreement between GYRE and the 3D simulation, confirming the type of the waves excited in the 3D simulations as being gravity waves. This is very satisfactory and allows us to advance further our analysis of their properties.

\subsection{Period spacing \label{sec:period_spacing}}

For $n \gg 1$, asymptotic $m=0$ g modes are evenly spaced in period.
The asymptotic period spacing for consecutive high-order g modes is \citep{1980ApJS...43..469T,Provost1986}:
\begin{equation}
    \label{eq:period_spacing}
    \overline {\Delta P}_\ell = \frac{\pi}{\sqrt{\ell(\ell+1)} \int_{r_1}^{r_2} \frac{N}{r} \mathrm{d} r}
,\end{equation}
where, for a mode of frequency $\omega$, $r_1$ and $r_2$ are the inner and outer boundary of the corresponding resonant cavity, respectively, defined by $N (r_1) = N (r_2) = \omega$.  

\begin{figure*}[ht]
    \centering
    \includegraphics[width=\textwidth]{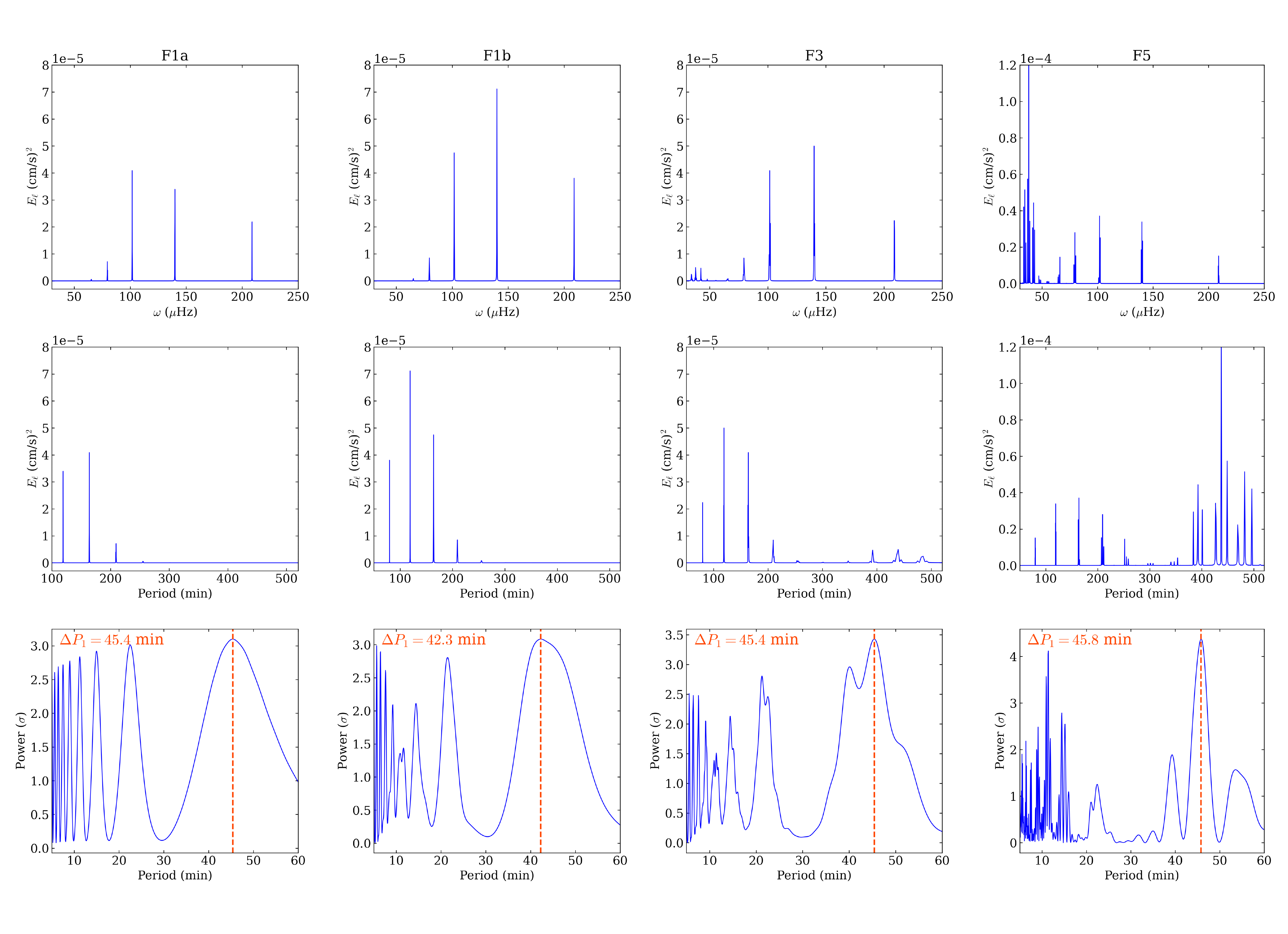}
    \caption{\textit{Top panels:} power spectra $E_1 \, (0.38 R_\star, \omega)$. \textit{Middle panels:} power spectra $E_1 \, (0.38 R_\star, P)$. \textit{Bottom panels:} Lomb-Scargle periodograms computed from $E_1 \, (0.38 R_\star, P)$. Each periodogram is normalised with its standard deviation, $\sigma$ and the obtained $\Delta P_l$ value is shown by a vertical dashed blue line. From left to right, the represented cases are F1a, F1b, F3, and F5.}
    \label{fig:periodogram_l1_d60}
\end{figure*}

\begin{figure*}[ht]
    \centering
    \includegraphics[width=\textwidth]{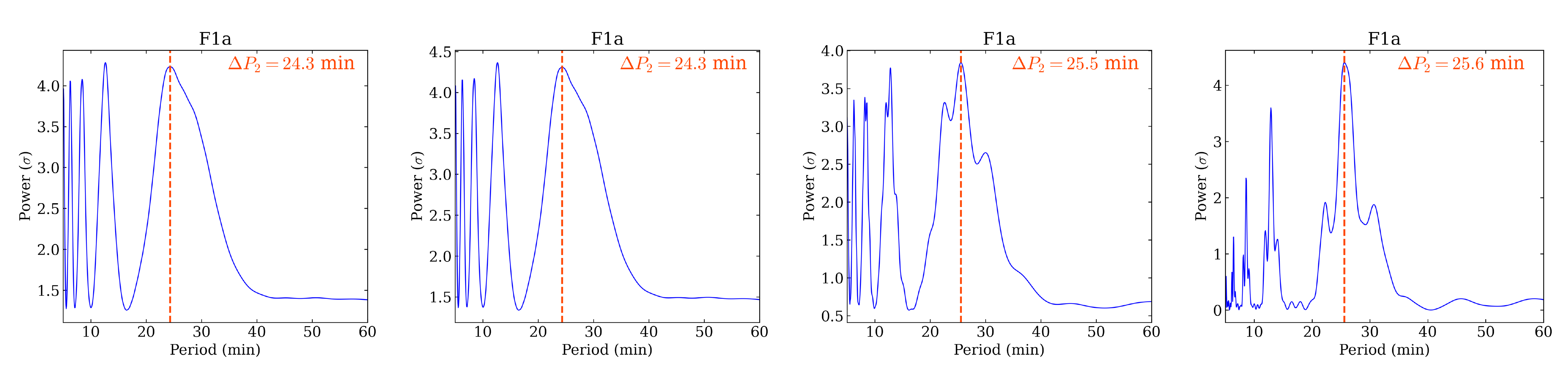}
    \caption{
    Lomb-Scargle periodograms computed from $E_2 \, (0.38 R_\star, P)$. Each periodogram is normalised with its standard deviation, $\sigma$ and the obtained $\Delta P_2$ value is shown by a vertical dashed red line. From left to right, the represented cases are F1a, F1b, F3, and F5.}
    \label{fig:periodogram_l2_d60}
\end{figure*}

In Fig.~\ref{fig:periodogram_l1_d60}, for each case, we represent the $E_1 \, (0.38 R_\star, \omega)$ power spectra and the corresponding $E_1 \, (0.38 R_\star, P = 1/\omega)$ power spectra scaled in period. The middle panels of Fig.~\ref{fig:periodogram_l1_d60} show the modes are almost equidistant in period.
Therefore, we consider 50 $< P <$ 520 min for $\ell=1$ and 50 $< P <$ 300 min for $\ell=2$, and, in order to highlight this $\Delta P_\ell$-periodicity in $E_\ell \, (0.38 R_\star, P)$, we use an approach similar to the one presented by \citet{2007Sci...316.1591G}, by computing the Lomb-Scargle periodogram \citep{1976Ap&SS..39..447L,1982ApJ...263..835S} of $E_\ell \, (0.38 R_\star, P$) for periods in the Lomb-Scargle periodogram spanning from 5 to 60 minutes. Each Lomb-Scargle periodograms is normalised by its standard deviation $\sigma$. We perform the same operation for the $E_2$ power spectra. The obtained Lomb-Scargle periodograms are shown in the bottom panels of Fig~\ref{fig:periodogram_l1_d60} for $E_1$ and in Fig.~\ref{fig:periodogram_l2_d60} for $E_2$.
We use the periodograms to identify the period spacings, $\Delta P_1$ and $\Delta P_2$, which we compare to the asymptotic value, $\overline {\Delta P}_1 = 47$ min and $\overline {\Delta P}_2 = 27.1$ min.
We find $\Delta P_1 = 45.4$, 42.3, 45.4, and 45.8 min and $\Delta P_1 = 24.3$, 25.8, 25.5, and 26 min for the F1a, F1b, F3, and F5 cases, respectively. These values are all below the asymptotic references, which is expected as we considered low-order modes for which period spacings are usually smaller than for higher order modes. It is nevertheless satisfactory that we are able to identify without ambiguity the periodicity related to the mode pattern in the Lomb-Scargle periodograms.
We also note that the shape of the peaks yielded by the periodogram significantly changes with rotation, as mode splittings is considerably more visible on the frequency range we consider for fast rotating models and the split components are not equidistant in period. 

The difference in amplitudes for low-frequency modes between slow-rotating cases F1a and F1b, and fast rotating cases F3 and F5 clearly appears in Fig.~\ref{fig:periodogram_l1_d60} and \ref{fig:periodogram_l2_d60}. On these two figures, we also note that high-frequency modes are less excited in the F3 and F5 cases than in the F1a and F1b, suggesting that the mode excitation efficiency is shifted towards low frequencies. 

We also compare the evolution of $\Delta P_\ell$ along the period with the outputs of GYRE computations. In order to take into account the Coriolis effect in our comparison, we use GYRE to compute the predicted mode frequencies for $\ell = 1 \; ; \; m = \{ -1, 0, 1 \}$ with the traditional approximation of rotation (TAR), assuming a solid body rotation at 5 $\Omega_\odot$. We compare the outputs with the F5 case.  
The TAR treatment is useful when the ratio $\omega / 2\Omega$ becomes to small to take rotation into account using only a perturbative approach \citep{1997ApJ...491..839L}. In the TAR, the Coriolis force is included in the equations to solve but its radial component is neglected. The TAR is therefore well-suited to study wave behaviour in stably stratified radiative interiors but breaks in convective zones. The GYRE run using the TAR will be referred in what follows as the 5 $\Omega_\odot$ GYRE TAR run.

In order to measure the period spacings in the simulation, we considered a 280-day long time series of the F5 case (see Table~\ref{tab:model_summary}) and we compute the corresponding $E_{\ell,m} (r = 0.13 R_\star, \omega)$ power spectrum (see Eq.~\ref{eq:ref_Elm}) to fit a Lorentzian profile for each $m$ component of the $\ell=1$ modes visible at this depth. The Lorentzian profile is fitted with a Markov chain Monte Carlo (MCMC) approach, assuming a $\chi^2$ statistics with two degrees of freedom, implemented with the \texttt{emcee} module \citep{2013PASP..125..306F}. 
We compute the corresponding periods, $P_{n,1,m}$, and period spacings, $\Delta P_{n,1,m}$, relative to the azimutal number, $m$, taken as: 
\begin{equation}
    \Delta P_{n,1,m} = P_{n+1,1,m} - P_{n,1,m} \ .
\end{equation}

\begin{figure}[ht!]
    \centering
    \includegraphics[width=.49\textwidth]{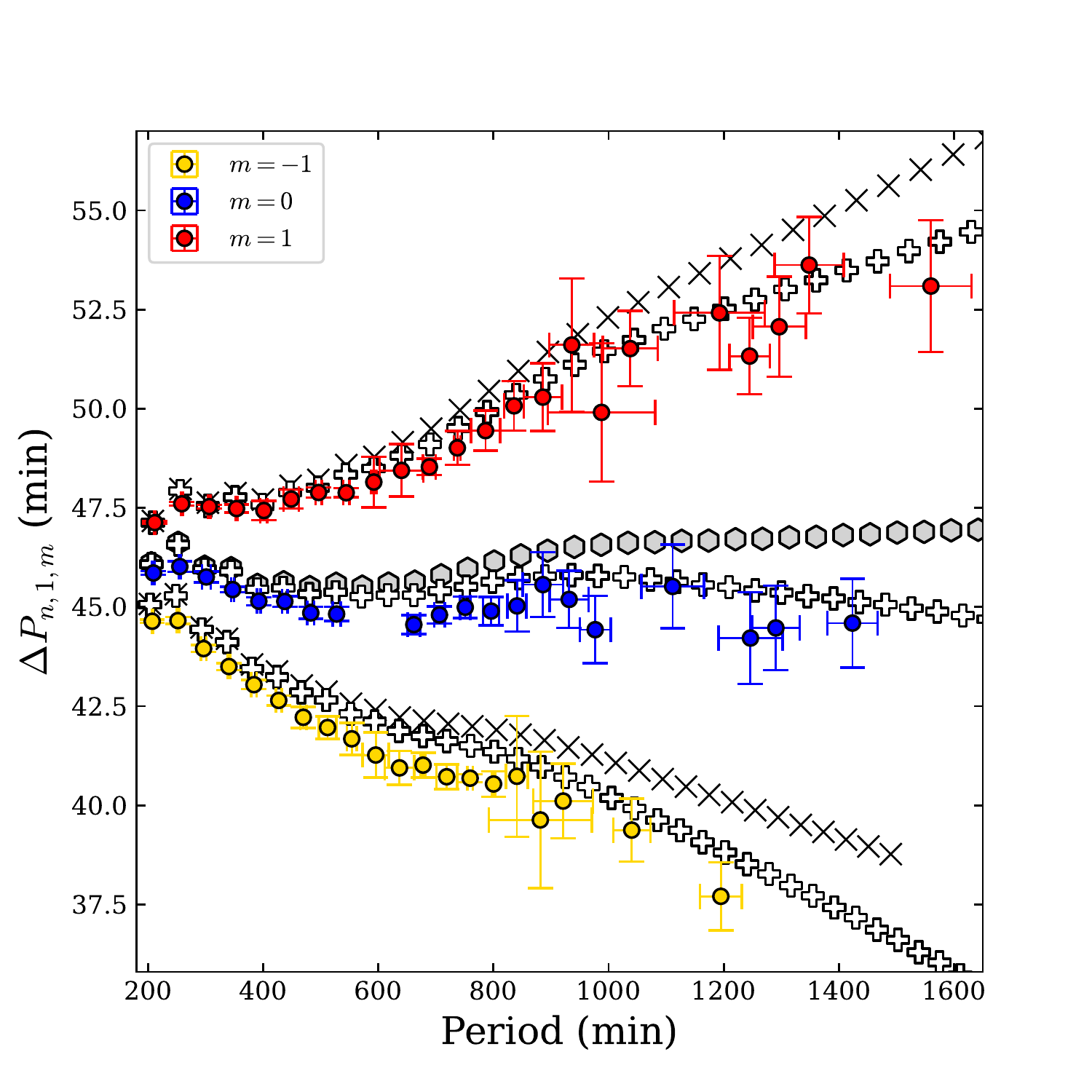}
    \caption{$\Delta P_1$ as a function of modes period in the F5 case, for azimuthal numbers $m = -1$ (\textit{yellow}), $m = 0$ (\textit{blue}), and $m = 1$ (\textit{red}). The $\Delta P_1$ obtained from the $\Omega_\star = 0$ GYRE run and the perturbative approach are represented as grey hexagons while the white crosses shows the $\Delta P_1$ obtained for $m = \{ -1, 0, 1 \}$ with the 5 $\Omega_\odot$ GYRE TAR run. The black crosses signal the $\Delta P_1$ position for $m = \{ -1, 1 \}$ in the asymptotic approximation of the perturbative method. 
    The measured uncertainties on period and $\Delta P_1$ are represented.
    We show only fitted modes for which we are able to measure $\Delta P_1$ with an uncertainty below 1.75 min.}
    \label{fig:delta_p_5_solar_rot_tar}
\end{figure}

The results of this analysis are shown in Fig.~\ref{fig:delta_p_5_solar_rot_tar}. We compare them with the $\Delta P_{n,1,m}$ obtained with the 5 $\Omega_\odot$ GYRE TAR run and the perturbative approach applied to the $\Omega_\star = 0$ GYRE runs. At long periods, the $\Delta P_{n,1,m}$, obtained considering the TAR significantly differs from what the perturbative approach predicts. Since the Coriolis force is fully taken into account in the equations solved by the ASH code, the $\Delta P_{n,1,m}$ evolution observed in the 3D simulation follows much better the 5 $\Omega_\odot$ GYRE TAR run. To assess this more quantitatively, we define $\mathcal{M}$, a least-square distance metric: 
\begin{equation}
    \mathcal{M} = \sum_{n,m} |\Delta P_{n, 1, m \, (\mathrm{ASH})} -  \Delta P_{n, 1, m \, (\mathrm{GYRE})} |^2 / \sigma_{n, 1, m}^2 \; , 
\end{equation}
where the $\sigma_{n, 1, m}^2$ are the uncertainties measured for each value of $\Delta P_{n, 1, m \, (\mathrm{ASH})} $.
As a least-square distance metric, $\mathcal{M}$ is equivalent to a log-likelihood with variables following a Gaussian distribution. 
Considering both approaches, we can therefore determine the most likely by computing $\delta = \mathcal{M}_\mathrm{p.a} - \mathcal{M}_\mathrm{TAR}$, where $\mathcal{M}_\mathrm{TAR}$ is the value obtained with the TAR and $\mathcal{M}_\mathrm{p.a}$ is the value obtained with the perturbative approach. A positive value of $\delta$ would favour the TAR hypothesis while a negative value would favour the perturbative approach hypothesis.
We obtain $\delta = 416$, which strongly favours the TAR hypothesis.
This provides another evidence that the period spacings that we measure in the simulation are more compatible with the values obtained using the TAR.

\subsection{Rotational splittings \label{sec:splittings}}

\begin{figure}[ht!]
    \centering
    \includegraphics[width=.49\textwidth]{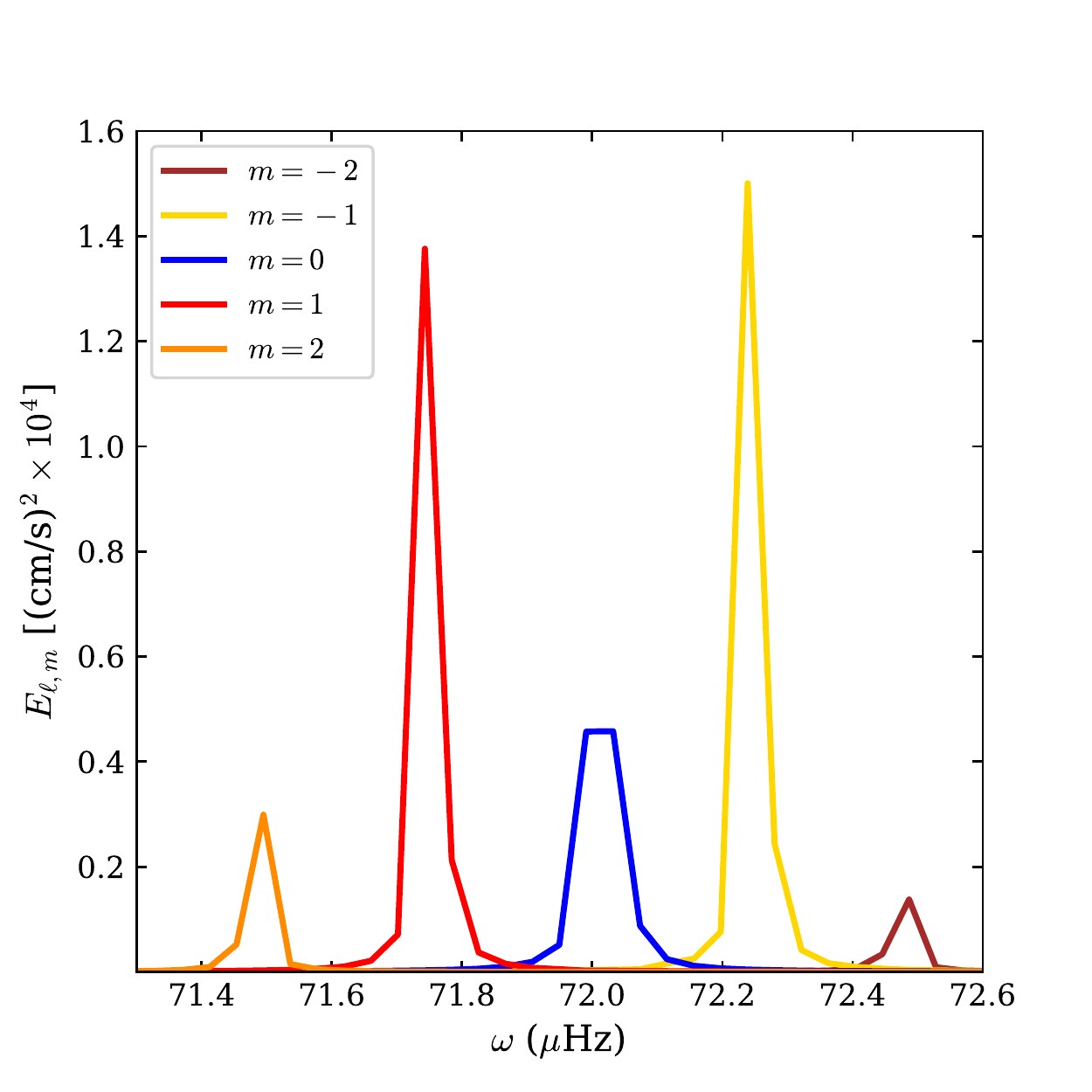}
    \caption{Example of rotational splitting observed in the F5 case for a $\ell=2$ mode, with the $m=-2$ shown in brown, the $m=-1$ in yellow, the $m=0$ in blue, the $m=1$ in red, and the $m=2$ in orange.}
    \label{fig:splitting_example}
\end{figure}

\begin{figure}[ht!]
    \centering
    \includegraphics[width=.49\textwidth]{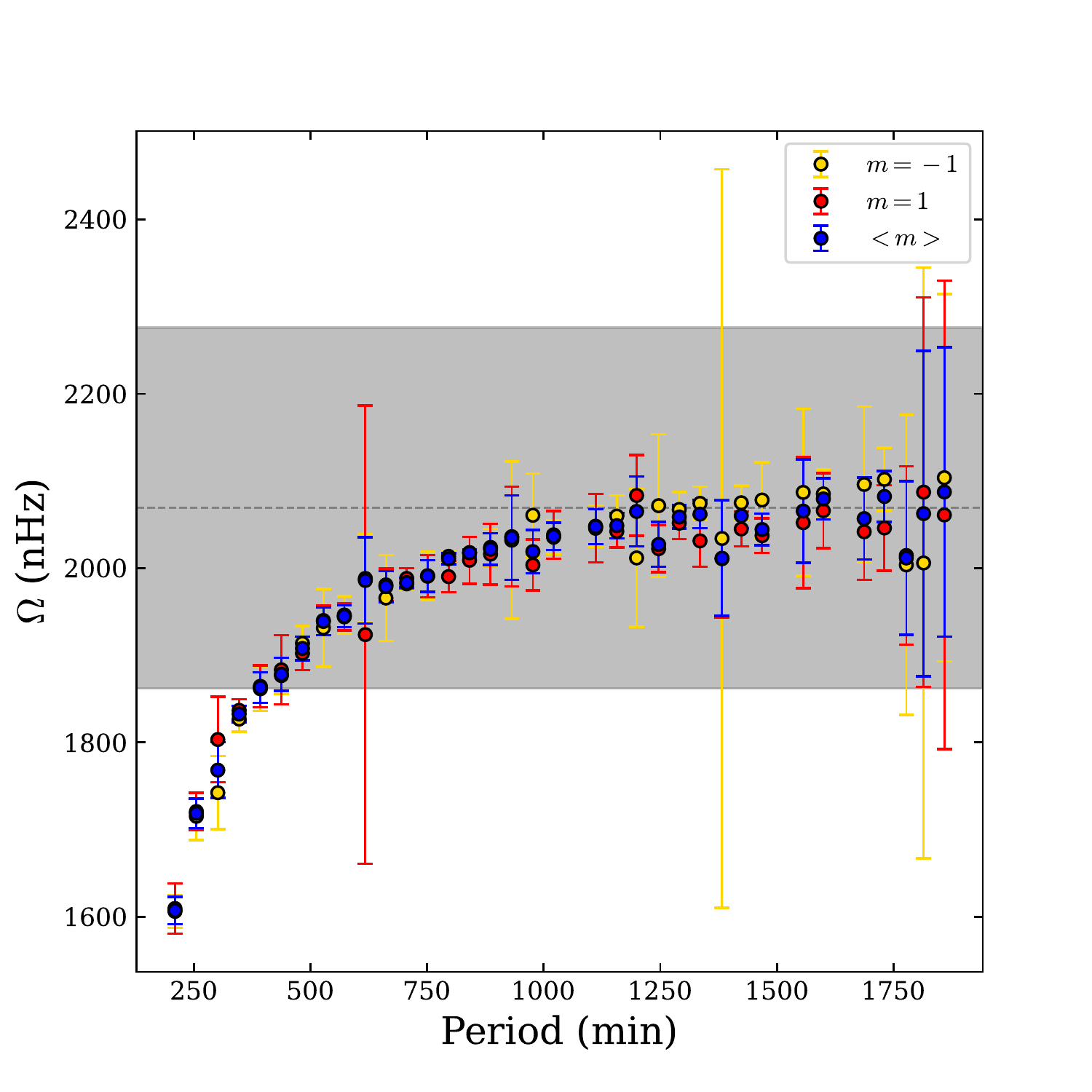}
    \caption{Rotation rate $\Omega$ inferred from the asymptotic approximation of Eq.~\ref{eq:delta_nlm_corotational_frame} and \ref{eq:beta_nl_approx} with $\delta_{n,1,m}$ measurements, in the F5 case. The position of the true rotation value of the model is highlighted by the dashed grey line, with the 10\% error interval in grey. Rotation rate measured only with $\delta_{n,1,-1}$ and $\delta_{n,1,1}$ are represented in yellow and red, respectively, while the averaged mean of the two values is shown in blue. The error bar for each $\Omega$ measurement is also represented.}
    \label{fig:splittings_5_solar_rot}
\end{figure}

In the case of uniform rotation and in the slow rotation limit, $\omega \gg 2 \Omega_0$, the rotational splitting for a mode component with frequency $\omega_{n\ell m}$ is given by \citep[e.g.][]{ChristensenDaalsgardLectureNotes}
\begin{equation}
\label{eq:splitted_frequency}
    \omega_{n\ell m} = \omega_{n\ell} + \delta_{n\ell m}
\end{equation}
with, in the inertial frame,  
\begin{equation}
\label{eq:delta_nlm_inertial_frame}
    \delta_{n\ell m} = m \beta_{n\ell} \Omega_0 \; ,
\end{equation}
and in the co-rotating frame,
\begin{equation}
\label{eq:delta_nlm_corotational_frame}
    \delta_{n\ell m} = - m (1 - \beta_{n\ell}) \Omega \; .
\end{equation}
The parameter $\beta_{n\ell}$ is: 
\begin{equation}
\label{eq:beta_nl}
    \beta_{n\ell} = \frac{\int_0^{R_\star} (\xi_r^2 + \mathcal{L}^2 \xi_h^2 - 2\xi_r\xi_h - \xi_h^2) r^2 \bar{\rho} \mathrm{d}r }{\int_0^{R_\star} (\xi_r^2 + \mathcal{L}^2 \xi_h^2 - 2\xi_r\xi_h) r^2 \bar{\rho} \mathrm{d}r} \; ,
\end{equation}
where $\xi_r$ and $\xi_h$ are the radial and horizontal displacements, respectively, and $\mathcal{L}=\sqrt{\ell (\ell+1)}$. In the case of high-order g modes, we can consider
\begin{equation}
\label{eq:beta_nl_approx}
    \beta_{n\ell} \approx 1 - \frac{1}{\ell (\ell + 1)} \; .
\end{equation}

Considering a $\ell=2$ mode, we provide an example of how rotational splittings appear in the power spectrum of the F5 case in Fig.~\ref{fig:splitting_example}. In the co-rotating frame, as expected from Eq.~\ref{eq:delta_nlm_corotational_frame}, the $m$-component frequency decreases as $m$ increase.
In order to measure the rotational splittings in the simulation, we use the mode frequencies measured in Sect~\ref{sec:period_spacing} for $\ell = 1$ modes.
For each fitted component $m$, we take the Lorentzian central frequency as the $m$ component frequency and we use Eq.~\ref{eq:delta_nlm_corotational_frame} and \ref{eq:beta_nl_approx} to compute the corresponding $\Omega$ value.
Figure~\ref{fig:splittings_5_solar_rot} shows the result of this analysis. 
We show the value inferred by using the $m=-1$ or the $m=1$ alone, along with the value inferred by considering the averaged mean of the two measurements.
We are able to fit almost every order with period $P$ below 1800 min. 
The one for which we do not provide a measurement are those who are close to a node ridge at this depth of the simulation and, therefore, we have therefore very low amplitude in the considered power spectrum. In most of the cases, the uncertainties over the frequency we measure are significantly below the 10\% interval around $\Omega_0$. As we are constrained by the resolution of the time series we use, we note that the uncertainties on the inferred $\Omega$ increase as we go towards long periods (low frequencies).
Obviously, the asymptotic approximation does not hold for modes of lowest orders and it is not possible to retrieve the correct rotation rate just by using the $\beta_{n\ell}$ approximation yielded by Eq.~\ref{eq:beta_nl_approx} with such low-$n$ modes.

\section{Mode tunneling and surface visibility \label{sec:surface_visibility}}

\begin{figure*}[ht!]
    \centering
    \includegraphics[width=\textwidth]{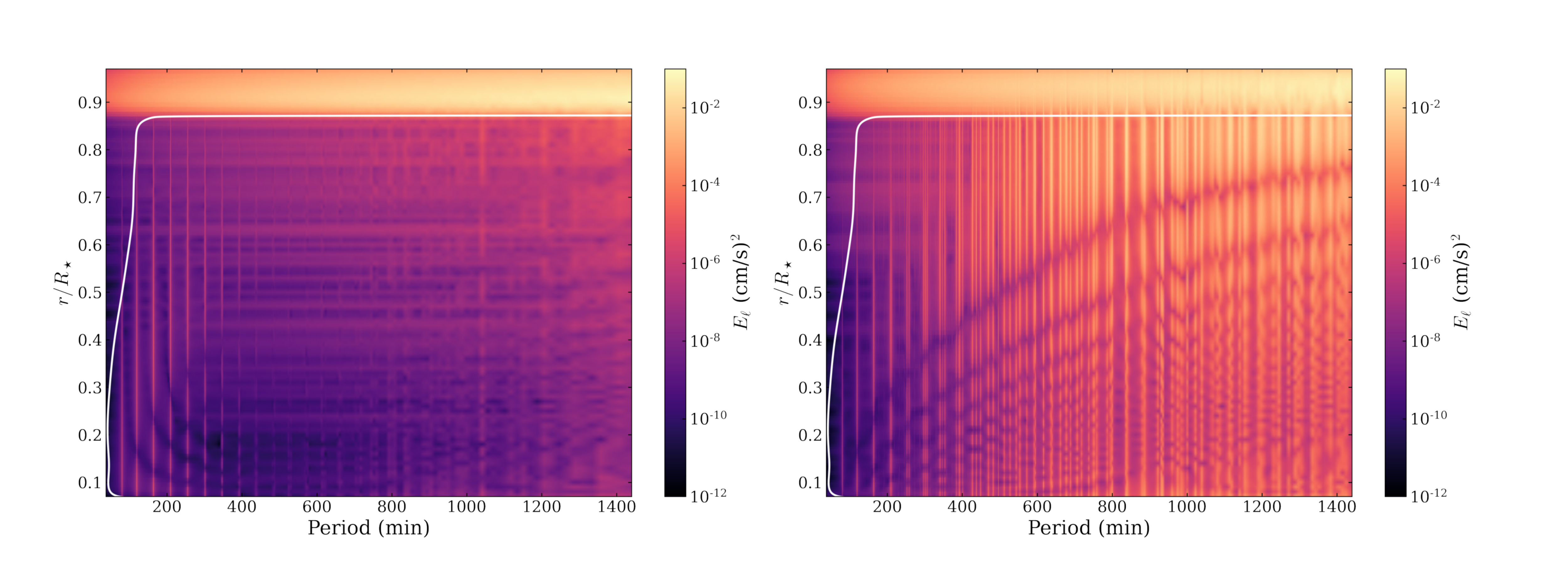}
    \caption{Power spectrum for $\ell=1$, $E_1$, radial evolution in the F1b (\textit{left}) and  F5 (\textit{right}) cases. 
    In each panel, the white line represents the boundary of the mode resonant cavity. The range of radial orders visible in the figure is $1 \leq n \leq 30$.}
    \label{fig:depth_map_l1}
\end{figure*}

\begin{figure*}[ht!]
    \centering
    \includegraphics[width=\textwidth]{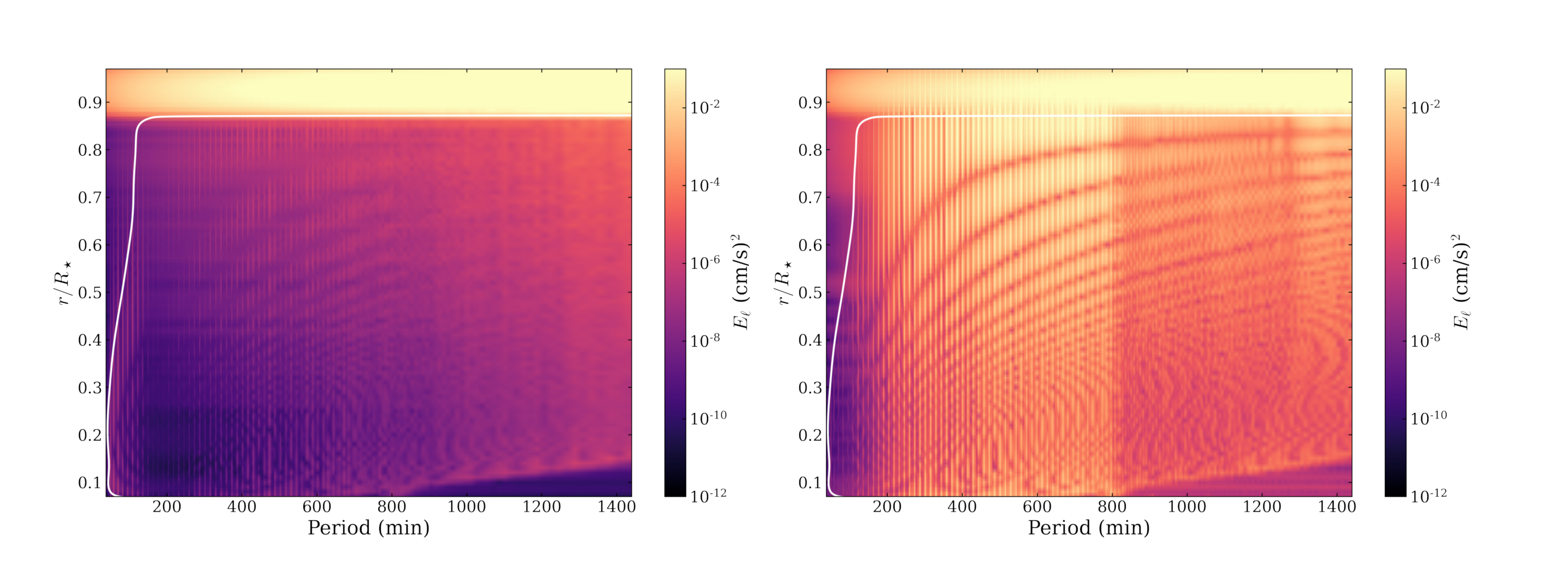}
    \caption{Power spectrum for $\ell=4$, $E_4$, radial evolution in the F1b (\textit{left}) and  F5 (\textit{right}) cases. 
    In each panel, the white line represents the boundary of the mode resonant cavity. The cutoff between g modes and progressive waves is visible around 800 min. Below this cutoff, the range of visible radial orders is $1 \leq n \leq 50$}
    \label{fig:depth_map_l4}
\end{figure*}

We go on to study the radial dependency of the mode amplitudes in the power spectrum. In Fig~\ref{fig:depth_map_l1} and \ref{fig:depth_map_l4}, we represent for F1b and F5 the radial evolution of $E_\ell$ for $\ell=1$ and $\ell=4$, respectively, and for $P < 1440$ min.  
The boundary between the resonant cavity and the evanescent region is represented by the white line. As expected, the power level is significantly higher in the convective envelope because of the contribution of convective motions. The location of the mode nodes draw dark ridges of low amplitudes in the spectrum. As already underlined in Sec.~\ref{sec:igw_spectrum}, the F3 and F5 cases exhibit more power in high-period (low-frequency) modes. 
The splitted character of the $\ell = 1$ modes at high period (low frequency) appears for the F3 and F5 cases. As expected from Eq.~\ref{eq:delta_nlm_corotational_frame}, in the co-rotational frame, mode splittings appear less clearly for $\ell = 4$ modes. 
For the $E_4$ power spectrum, the separation between g modes and progressive waves is visible in Fig.~\ref{fig:depth_map_l4} and located close to $P = 800$ min. Therefore, modes with a radial order below $n=50$ are visible in the figure, while the range presented for the $\ell = 1$ modes in Fig.~\ref{fig:depth_map_l1} covers $1 \leq n \leq 30$.
As underlined in Sect.~\ref{sec:igw_spectrum} (see Fig.~\ref{fig:power injection}), modes with $\ell=4$ are particularly excited in the F5 case. The bottom right panel of Fig.~\ref{fig:depth_map_l4} shows that, for $200 < P < 800$~min and above $r = 0.88 \, R_\star$, the mode signature is clearly visible above the white line and among the convective flows contribution.

\begin{figure}[ht]
    \centering
    \includegraphics[width=0.49\textwidth]{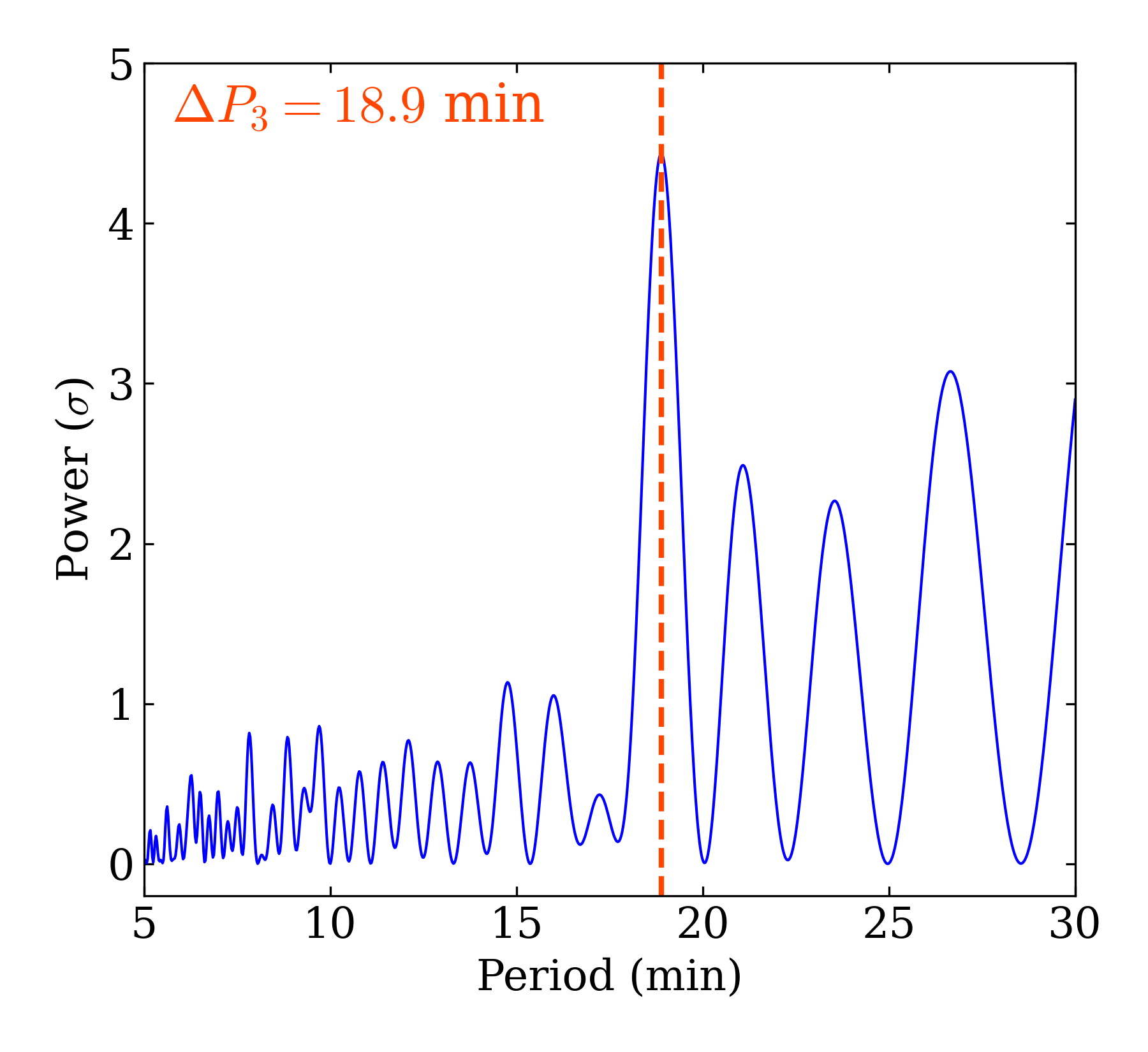}
    \includegraphics[width=0.49\textwidth]{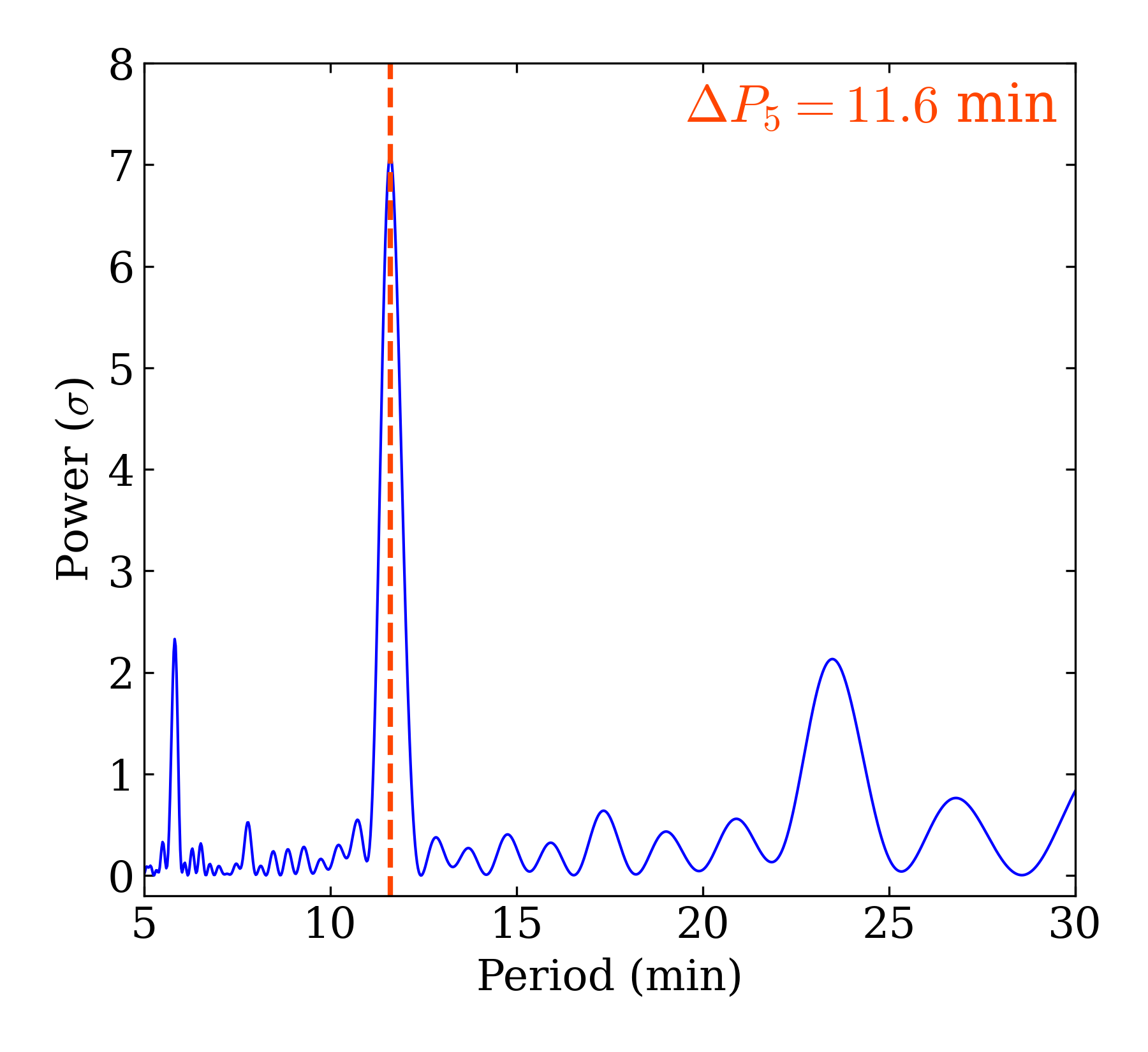}
    \caption{Lomb-Scargle periodograms computed from $E_3 \, (0.97 R_\star, P)$ (\textit{top}) and $E_5 \, (0.97 R_\star, P)$ (\textit{bottom}). The periodogram is normalised with its standard deviation $\sigma$ and the obtained $\Delta P_3$ value is shown by a vertical dashed red line.}
    \label{fig:l3_l5_signature_top_domain}
\end{figure}

As illustrated with $\ell=4$ modes, the F5 case presents evidence of low- and mid-degree modes tunneling through the convective zone and therefore we look for g-mode signatures in the convective velocity signal near the top of the domain, where we still have $\tilde{v}_r$ of the order of \num{1.5e4} cm/s (see Table~\ref{fig:vrms_shav}). The F5 case is the only model where clear g-mode patterns are observable in the convective envelope. 
Using the same periodogram method presented in Sec.~\ref{sec:period_spacing}, we look for periodicity in the $v_r$ signal at $r = 0.97$ $R_\star$ for $1 \leq \ell \leq 10$, considering the period range 250 to 450 min. 
We are able to detect the g-mode signature of $\ell=3$, 4, 5, 6, and 7 modes. 
This is particularly interesting for $\ell = 3$ modes which are still observable in disk-integrated observations. 
As an illustration, we show in Fig.~\ref{fig:l3_l5_signature_top_domain} the Lomb-Scargle periodograms computed for $\ell=3$ and $\ell=5$ modes, for the F5 case. 
The clearest detection is the $\Delta P_5$, with a peak height at 7.1 $\sigma$ against 4.4 $\sigma$ for $\Delta P_3$, and 5.8 for $\Delta P_4$. We report in Table~\ref{tab:delta_p_top_domain} the measured $\Delta P_\ell$ and corresponding peak heights. They are compared with the asymptotic $\overline{\Delta P}_\ell$ computed with Eq.~\ref{eq:period_spacing}. 

\begin{table}[ht!]
    \centering
    \caption{Asymptotic $\overline{\Delta P}_\ell$, $\Delta P_\ell$ measured at  $r = 0.97$ $R_\star$ and corresponding peak heights for the F5 case.}
    \begin{tabular}{cccc}
    \hline
    \hline
        $\ell$ & $\overline{\Delta P}_\ell$ (min) & $\Delta P_\ell$ (min) & Peak height ($\sigma$) \\
    \hline
            3 & 19.2 & 18.9 & 4.4 \\
            4 & 14.9 & 14.3 & 5,8 \\
            5 & 12.1 & 11.6 & 7.1 \\
            6 & 10.3 & 9.8  & 5.2 \\
            7 & 8.9  & 8.5  & 5.3 \\
    \end{tabular}
    \label{tab:delta_p_top_domain}
\end{table}

Finally, we estimate the bolometric luminosity perturbation for individual g modes. In what follows, all the considered perturbations, $\delta L_\star$, $\delta R_\star$, and $\delta T_\mathrm{eff}$ are related to the action of an individual mode. 
From the Stefan-Boltzmann law we have \citep[e.g.][]{2010Ap&SS.328..253S}
\begin{equation}
    \frac{\delta L_\star}{L_\star} = 4 \frac{\delta T_\mathrm{eff}}{T_\mathrm{eff}} + 2 \frac{\delta R_\star}{R_\star} \; ,
\end{equation}
and, following \citet{Townsend2003b}
\begin{equation}
    \frac{\delta T_\mathrm{eff}}{T_\mathrm{eff}} = \nabla_\mathrm{ad} \left[ \frac{\ell (\ell+1)}{\overline{\omega}^2} - 4 - \overline{\omega}^2 \right] \; ,
\end{equation}
where $\nabla_\mathrm{ad} = 0.4$ is the adiabatic temperature gradient, and $\overline{\omega}^2$ is a dimensionless frequency defined as
\begin{equation}
    \overline{\omega}^2 = \frac{\omega^2 R_\star^3}{G M_\star} \; .
\end{equation}

To compute the displacement $\delta R_\star$ related to a single mode, we simply consider:
\begin{equation}
   \delta R_\star = \frac{\tilde{v}_{r, \mathrm{mode}}}{2 \pi \omega} \; ,
\end{equation}
where $\tilde{v}_{r, \mathrm{mode}}$ is the radial rms velocity (measured in the $E_\ell$ power spectrum) of the considered mode. 

\begin{figure}[ht]
    \centering
    \includegraphics[width=0.49\textwidth]{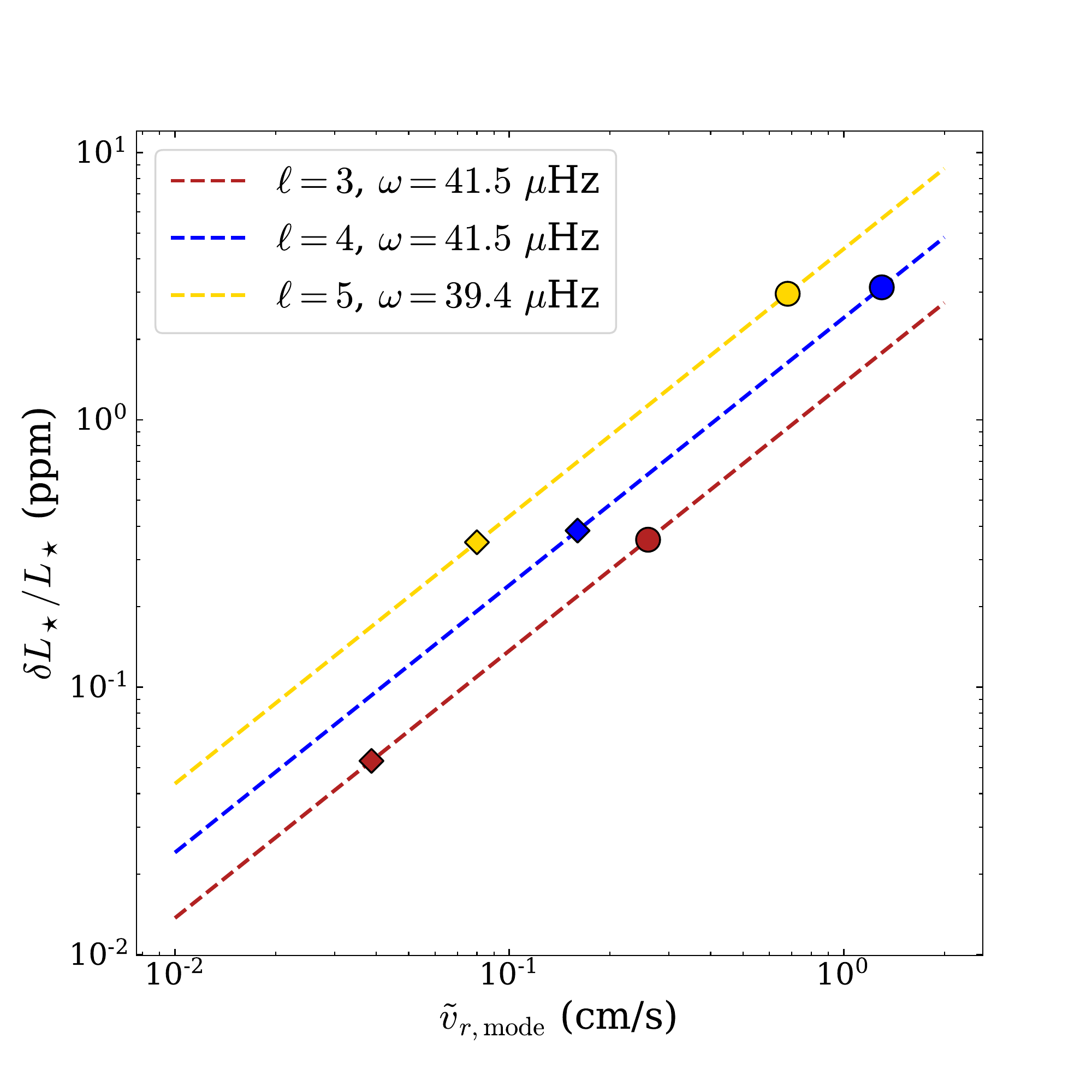}
    \caption{Luminosity perturbation estimation as a function of $v_r$ for a $\ell=3$ mode at 41.5 $\mu$Hz (\textit{red}), a $\ell=4$ mode at 41.5 $\mu$Hz (\textit{blue}), and a $\ell=5$ mode at 39.4 $\mu$Hz (\textit{yellow}). The diamonds correspond to the radial velocity we measure in the F5 case at $r=0.87$ $R_\star$ and the dots to the same quantity at $r=0.97$ $R_\star$.}
    \label{fig:luminosity_perturbation}
\end{figure}

For $\ell = 3$, 4, and 5, that is the degrees with the most excited individual modes, we compute the luminosity perturbation $\delta L_\star / L_\star$ as a function of $\tilde{v}_{r, \mathrm{mode}}$ for the modes with the largest amplitude at the top of the radiative zone. These three modes happen to have close eigenfrequencies, 41.5, 41.5, and 39.4 $\mu$Hz, respectively, and correspond to $n=19$, 23, and 28.   
We consider both the mode radial velocity at the top of the radiative zone, $r = 0.87$ $R_\star$, and near the top of the domain, $r = 0.97$ $R_\star$.
The results of this analysis are represented in Fig.~\ref{fig:luminosity_perturbation}. When using the estimate at the top of the radiative zone, we find corresponding $\delta L_\star / L_\star$ of a few part-per-millions (ppm) for the $\ell=4$ and $\ell=5$ mode. The value for the $\ell=3$ mode is significantly smaller, close to 0.3 ppm. To maintain a propagative behaviour above the upper limit of the Brunt-Väisala resonant cavity, gravito-inertial waves related to the modes would have to be sub-inertial \citep{2014A&A...565A..47M}, which is not the case here, as for the F5 case, we have $2\Omega_\star = 4.138$ $\mu$Hz.
Nevertheless, as their eigenfrequencies are close to $10\Omega_\star$ where the inertial contribution still has a significant influence on the wave behaviour (Fig.~\ref{fig:delta_p_5_solar_rot_tar} and Mathis, private communication), it is likely that the super-inertial character of these modes increases the characteristic length of their evanescent tail in the convective envelope, compared to slower rotating case. The luminosity perturbation for mode velocities near the top of the domain are all below 1 ppm. 
The $\delta L_\star / L_\star$ estimated in our analysis are several orders of magnitudes below what is typically observed for unstable modes in $\gamma$ Doradus stars. This underlines that the stochastic excitation of stable modes remains much less efficient than unstable mechanisms. We remind that our model has been chosen to be compatible with the existence of solar-type p-mode oscillations, but has $T_\mathrm{eff}$ and $\log g$ values that locate it close to the $\gamma$ Doradus instability strip lower boundary. However, evidence were presented above that g-mode may efficiently tunnel through the convective zone and may be detected through their $\Delta P_\ell$ signature. This leaves open the perspective to detect stochastically excited g modes and gravito-inertial modes in late F-type stars.

\section{Discussion and conclusion \label{section:conclusion}}

In this work, we presented and studied the first deep-shell 3D simulation of IGWs excited in the radiative interior of a F-type solar-type star by convective motions. We considered a 1.3 $M_\odot$ model for which we ran simulations at 1, 3, and 5 $\Omega_\odot$, which are rotation rates representative of F-type stars surface rotation rates observed by \textit{Kepler} (see Fig.~\ref{fig:santos_2021_distribution}). We described the effect of rotation on the convective structure and on the differential rotation regime in the convective zone. The properties of the stochastically excited IGWs were extensively studied. In particular, we showed that the excitation rate of low-frequency waves was significantly higher for fast rotating cases. We compared the eigenfrequencies of the modes in the simulations with the result of computations performed with the 1D oscillation code GYRE and we found a good agreement between the two approaches. 
We verified that the internal rotation rate could be inferred from rotational splittings in the asymptotic regime. Taking advantage of the regular period spacing of g modes, we computed the Lomb-Scargle periodograms of the power spectra to detect this periodicity and we compared the measured values with asymptotic predictions. Studying the period-spacing evolution at long periods in the F5 case, we were also able to observe the effect of the Coriolis force on the behaviour of high-order modes. Finally, we tried to quantify the possibility of observing g-mode signatures in actual F-type solar-type pulsators. We showed (see Fig.~\ref{fig:l3_l5_signature_top_domain} and Table~\ref{tab:delta_p_top_domain}) that for intermediate degrees $\ell=3$, 4, 5, 6, and 7, the mode signature could still be detected near the top of the simulation domain. As highlighted by Fig.~\ref{fig:luminosity_perturbation}, the most efficiently excited modes have a surface velocity of $\sim 10^{-1}$ cm/s, corresponding to a bolometric luminosity perturbation of $\sim 10^{-1}$ ppm, for the frequency range of these modes ($\sim 40$ $\mu$Hz). 
It should be underlined that these values are much larger than what was observed in the solar models from A14, where the maximal surface velocity (obtained with the most turbulent model) was $\sim \num{1e-6}$ cm/s. 
Near the top of the radiative zone (before tunneling through the evanescent region), we measure some mode velocities above 1 cm/s in the F5 case, corresponding to a luminosity perturbation of $\sim$3~ppm.
In the F1a and F1b cases, considering the most excited modes, we find near the top of the radiative zone typical velocities  of $\sim\num{1e-3}$ cm/s which is more than one order of magnitude higher than the maximal value observed by A14 near the tachocline ($\sim \num{5e-5}$ cm/s). These differences between the solar model and our F-type star models at $1 \; \Omega_\odot$ (F1a and F1b) can be explained by considering that the power injection from convection to the waves find a kinetic energy flux proportional to $\rho_b v_b^4$ \citep[see e.g.][]{Press1981,2016A&A...588A.122P} where $\rho_b$ is the density at the base of the convective zone and $v_b$ is a characteristic velocity at the base of the convective zone. For a given rotation rate, we therefore expect the rms velocity of the modes, $\tilde{v}_\mathrm{mode}$ to be subject to the following scaling formula:
\begin{equation}
\tilde{v}_\mathrm{mode} \propto \left (\frac{ \rho_b v_b^4} { \rho_{b,\odot} v_{b,\odot}^4 } \right)^{1/2} \; ,
\end{equation}
where the $\odot$ index denotes the solar values.
By comparing F1a and F1b with the solar model from A14, we find on one hand that convective motions at the bottom of the convective zone are significantly faster for our F-type model, by a factor of approximately 20 for the rms velocity, and 10 for the maximal velocity of the downward flows. On the other hand, the density at the interface is $\sim$100 times lower in the F-type case, thus meaning that the rms velocity of the modes should be $\sim10$ to 40 times larger, which is roughly consistent with what we see in the simulations. This considered, we therefore strongly advocate for the realisation of dedicated parametric studies of plume properties at the interface between the convective envelope and the radiative zone, including the effect of rotation. Such simulations would help to constrain the analytic models of IGWs excitation by convection, and would allow predicting the mode surface amplitude for a large range of models.
Indeed, the inclusion of rotation in our work shows that a parameterisation from 3D simulations of the form: 
\begin{equation}
\tilde{v}_\mathrm{mode} \propto \left (\frac{ \rho_b v^4 } { \rho_{b,\odot} v_{b,\odot}^4 }\right)^{1/2} \left (\frac{Ro}{\rm Ro_\odot} \right)^{a} \; ,
\end{equation}
with $a$ an unknown scaling parameter, could be compared with the theoretical predictions from \citet{2020ApJ...903...90A} and would prove extremely useful for future studies of IGWs in rotating stars.

We underline that some caveat must be kept in mind when considering these simulations. Due to the numerical constraints, we remind that the fluid regimes that are considered here are far from the actual regime in stellar interiors. The convective envelopes are in reality significantly more turbulent than what is currently achieved in any 2D or 3D simulation. 
In order to model a radiative zone as realistic as possible, we took $\kappa$ and $\nu$ values that are five and four orders of magnitudes lower to the values used in the convective zone.
In actual stellar interiors with $Pr$ thought to be comprised between \num{1e-5} and \num{1e-9} \citep[e.g.][]{Garaud2021}, radiative diffusion dominates the damping of the waves as they travel through the radiative regions, in a quasi-adiabatic regime \citep{1997A&A...322..320Z}, which should allow modes to form below a much lower frequency cutoff than observed in our simulations.
The $\tilde{v}_r$ values in the convective zone are constrained by $L_\star$. Therefore, it is mainly the profile of the power transfer function from convection to the wave and the ability of the waves to propagate through the radiative interior to form standing modes that will determine the shape of the g-mode power spectrum.

This work opens some observational perspectives for F-type solar-type-pulsating stars. For our fastest rotating case, F5, we were able to detect the signature of $\ell=3$ modes at the top of the domain, a spherical harmonic degree which is still accessible to observations from disk-integrated instruments like the \textit{Kepler} satellite, the Transiting Exoplanet Survey Satellite \citep[TESS,][]{Ricker2015}, or the PLAnetary Transits and Oscillations of stars \citep[PLATO,][]{Rauer2014} satellite. 
For targets bright enough to consider this type of follow-up, ground observations with échelle spectrograph as the ones from the Stellar Observations Network Group \citep[SONG,][]{Grundahl2007} could also be of greatest interest in the perspective of improving the characterisation of low-frequency regions of the brightest main-sequence F-type stars with solar-type oscillations observed by \textit{Kepler}.  
As our simulations show that $\ell=4$, 5, 6, and 7 modes are also particularly excited, long-term observations of some F-type stars from the \textit{Kepler} LEGACY sample \citep{2017ApJ...835..172L,SilvaAguirre2017}, with a stellar imager instrument dedicated to asteroseismology represent an interesting perspective \citep[e.g.][]{Christensen-Dalsgaard2011}. The estimated luminosity perturbations induced by an individual mode remains small (below 1 ppm) from what we compute with the mode velocity near the top of the domain, especially with the fact that we deal with modes laying in frequency regions where the power contribution from the convective signal is important. However, this suggests that modes excited by a more efficient mechanism like tidal forcing \citep[e.g.][]{Fuller2017} could reach the surface with an amplitude large enough to be detectable in these stars. 
In the future, we plan to include the convective core in the simulated domain in order to study the behaviour of IGWs excited simultaneously at the internal and the external interfaces of the radiative zone. The convective motions inside the core are also susceptible to play a significant role from a magneto-hydrodynamic point of view.
Indeed, in this work, we did not take the effect of the magnetic field into account. \citet{Lecoanet2022} recently showed that high-order g modes can be suppressed through the action of a strong internal magnetic field generated by convective motions in the core.

\begin{acknowledgements}
The authors thank the referee for useful comments that helped improve the paper. S.N.B., A.S.B and R.A.G acknowledge the support from PLATO CNES grant. S.N.B. and A.S.B acknowledge the support from Solar Orbiter CNES grant and financial support by ERC Whole Sun Synergy grant \#810218. S.N.B and R.A.G acknowledge the support from GOLF CNES grant. S.N.B reserves a special acknowledgment to L.~Amard for his precious knowledge in stellar structure and evolution and is also grateful to Q.~Noraz and A.~Strugarek for advice relative to the ASH code and its outputs. He also thanks F.~Grundahl for insights concerning the capabilities of  the SONG telescopes. Finally, the authors thank K.~Augustson, M.~Delorme, A. Le Saux, S.~Mathis, and C.~Pinçon for fruitful discussions.
\\
\textit{Software:} ASH \citep{CLUNE1999361,2004ApJ...614.1073B}, \texttt{Python} \citep{10.5555/1593511}, \texttt{numpy} \citep{numpy,Harris_2020}, \texttt{matplotlib} \citep{4160265}, \texttt{scipy} \citep{2020SciPy-NMeth}, \texttt{emcee} \citep{2013PASP..125..306F}.
\\
\end{acknowledgements}

\bibliographystyle{aa} 
\bibliography{biblio.bib} 

\appendix

\section{Spherical harmonic expansion and power spectrum computation \label{appendix:sph_expansion}}

This section provides some practical details on the techniques we used to obtain the power spectra $E_\ell$. In practice, we only compute the development for $m \geq 0$. We then consider the Fourier transform  $\hat{v_r} (r, \ell, m, \omega)$ of the complex quantity $v_r(r, \ell, m, t)$, with $- \omega_N < \omega < \omega_N$. From the spherical harmonic properties, we have $\hat{v_r} (r, \ell, -m, \omega) = \hat{v_r} (r, \ell, m, -\omega)$. The power contribution of the retrogade modes, $m<0$, therefore lies in $- \omega_N < \omega < 0$. The power contribution of the prograde modes, $m>0$ is in $0 < \omega < \omega_N$. The power contribution of a zonal mode, $m=0$, at frequency $\omega$, is split between $\omega$ and $-\omega$. For $\omega \geq 0$, the power in each $\ell,m$ component is: 

\begin{equation}
\label{eq:E_lm}
\left \{
\begin{aligned}
    E_{\ell,m} (r, \omega) &= |\hat{v_r}(r, \ell, m, \omega)|^2 \; ; \; &m>0, \; \\
    E_{\ell,m} (r, \omega) &= |\hat{v_r}(r, \ell, -m, -\omega)|^2 \; ; \; &m<0, \; \\
    E_{\ell,0} (r, \omega) &= |\hat{v_r}(r, \ell, 0, \omega)|^2 + |\hat{v_r}(r, \ell, 0, -\omega)|^2 \;  \;   .
\end{aligned}
\right .
\end{equation}

To compute the power spectrum $E_\ell (r, \ell, \omega)$ and restrict the domain to the positive frequencies $\omega \geq 0$, we consider for each degree $\ell$
\begin{equation}
\label{eq:E_l}
    E_\ell (r, \omega) = \sum_{m=-\ell}^{\ell} E_{\ell,m} (r, \omega) \; ; \; \omega \geq 0 \;  .
\end{equation}

\end{document}